\documentclass[iop]{emulateapj}

\usepackage{color}
\usepackage{amssymb, amsmath}

\usepackage[utf8]{inputenc}
\usepackage{lipsum}
\usepackage{pifont}
\usepackage{bm}
\usepackage{ulem}
\usepackage{balance}
\usepackage{multirow}
\usepackage{mathtools}
\usepackage{enumitem}
\usepackage{booktabs}
\usepackage{appendix}
\usepackage{listings}

\shorttitle{The \textsc{Quijote} simulations}
\shortauthors{Francisco Villaescusa-Navarro et al.}

\usepackage[usenames,dvipsnames]{xcolor}
\usepackage{hyperref}
\hypersetup{
    colorlinks = true,
    citecolor = {MidnightBlue},
    linkcolor = {BrickRed},
    urlcolor = {blue}
}

\newcommand{\be}{\begin{equation}}
\newcommand{\ee}{\end{equation}}
\newcommand{\ba}{\begin{eqnarray}}
\newcommand{\ea}{\end{eqnarray}}

\usepackage{soul}
\definecolor{nicegreen}{HTML}{2CA02C}
\definecolor{orange}{rgb}{1,0.5,0}

\begin{document}

\title{The Quijote simulations}

\author{Francisco Villaescusa-Navarro$^{1,2,\dagger}$, ChangHoon Hahn$^{3,4}$, Elena Massara$^{1,5}$, Arka Banerjee$^{6,7,8}$, Ana Maria Delgado$^{9,1}$, Doogesh Kodi Ramanah$^{10,11}$, Tom Charnock$^{10}$, Elena Giusarma$^{1,12}$, Yin Li$^{1,3,4,13,31}$, Erwan Allys$^{14}$, Antoine Brochard$^{15,16}$, Cora Uhlemann$^{17,18}$, Chi-Ting Chiang$^{19}$, Siyu He$^1$, Alice Pisani$^2$, Andrej Obuljen$^{5}$, Yu Feng$^{3,4}$, Emanuele Castorina$^{3,4}$, Gabriella Contardo$^1$, Christina D. Kreisch$^2$, Andrina Nicola$^2$, Justin Alsing$^{20,1}$, Roman Scoccimarro$^{21}$, Licia Verde$^{22,23}$, Matteo Viel$^{24,25,26,27}$, Shirley Ho$^{1,2,28}$, Stephane Mallat$^{29,30}$, Benjamin Wandelt$^{10,11,1}$, David N. Spergel$^{2,1}$}

\affil{$^1$Center for Computational Astrophysics, Flatiron Institute, 162 5th Avenue, 10010, New York, NY, USA}
\affil{$^2$ Department of Astrophysical Sciences, Princeton University, Peyton Hall, Princeton NJ 08544-0010, USA}
\affil{$^3$ Department of Physics, University of California, Berkeley, CA 94720, USA}
\affil{$^4$ Berkeley Center for Cosmological Physics, Berkeley, CA 94720, USA}
\affil{$^5$ Waterloo Centre for Astrophysics, University of Waterloo, 200 University Ave W, Waterloo, ON N2L 3G1, Canada}

\affil{$^6$ Kavli Institute for Particle Astrophysics and Cosmology, Stanford University, 452 LomitaMall, Stanford, CA 94305, USA}
\affil{$^7$ Department of Physics, Stanford University, 382 Via Pueblo Mall, Stanford, CA 94305, USA}
\affil{$^8$ SLAC National Accelerator Laboratory, 2575 Sand Hill Road, Menlo Park, CA 94025, USA}

\affil{$^9$ Department of Physics, New York City College of Technology, Brooklyn, NY 11201, USA}

\affil{$^{10}$ Sorbonne Universite, CNRS, UMR 7095, Institut d'Astrophysique de Paris, 98 bis boulevard Arago, 75014 Paris, France}
\affil{$^{11}$ Sorbonne Universite, Institut Lagrange de Paris, 98 bis boulevard Arago, 75014 Paris, France}

\affil{$^{12}$ Department of Physics, Michigan Technological University, Houghton, MI, 49931, USA.}

\affil{$^{13}$ Kavli Institute for the Physics and Mathematics of the Universe (WPI), UTIAS, The University of Tokyo, Chiba 277--8583, Japan}

\affil{$^{14}$ Laboratoire de Physique de l'Ecole normale superieure, ENS, Universite PSL, CNRS, Paris, France}
\affil{$^{15}$ INRIA, ENS, PSL Research University Paris, France}
\affil{$^{16}$ Paris Research Center, Huawei Technologies, Paris, France}

\affil{$^{17}$ Centre for Theoretical Cosmology, DAMTP, University of Cambridge, CB3 0WA Cambridge, United Kingdom}
\affil{$^{18}$ Fitzwilliam College, University of Cambridge, CB3 0DG Cambridge, United Kingdom}

\affil{$^{19}$ Physics Department, Brookhaven National Laboratory, Upton, NY 11973, USA}

\affil{$^{20}$ Oskar Klein Centre for Cosmoparticle Physics, Department of Physics, Stockholm University, Stockholm SE-106 91, Sweden}

\affil{$^{21}$ Center for Cosmology and Particle Physics, Department of Physics, New York University, NY 10003, New York, USA}

\affil{$^{22}$ Institut de Ciencies del Cosmos, University of Barcelona, ICCUB, Barcelona 08028, Spain}
\affil{$^{23}$ Institucio Catalana de Recerca i Estudis Avancats, Passeig Lluis Companys 23, Barcelona 08010, Spain}

\affil{$^{24}$ SISSA, Via Bonomea 265, 34136 Trieste, Italy}
\affil{$^{25}$ INFN, Sez.  di Trieste, Via Valerio 2, 34127 Trieste, Italy}
\affil{$^{26}$ IFPU, Institute for Fundamental Physics of the Universe, via Beirut 2, 34151 Trieste, Italy}
\affil{$^{27}$ INAF, Osservatorio Astronomico di Trieste, via Tiepolo 11, I-34131 Trieste, Italy}

\affil{$^{28}$Department of Physics, Carnegie Mellon University, Pittsburgh, PA 15213, USA}

\affil{$^{29}$ Data team, Ecole Normale Sup\'erieure, Universit\'e PSL, 45 rue d'Ulm, 75005 Paris, France}
\affil{$^{30}$ College de France, 11 place Marcelin Berthelot, 75005, Paris, France}
\affil{$^{31}$ Center for Computational Mathematics, Flatiron Institute, 162 5th Avenue, 10010, New York, NY, USA}

\altaffiltext{$\dagger$}{villaescusa.francisco@gmail.com}

\begin{abstract}
The \textsc{Quijote} simulations are a set of 44,100 full N-body simulations spanning more than 7,000 cosmological models in the $\{\Omega_{\rm m}, \Omega_{\rm b}, h, n_s, \sigma_8, M_\nu, w \}$ hyperplane. At a single redshift the simulations contain more than 8.5 trillions of particles over a combined volume of 44,100 $(h^{-1}{\rm Gpc})^3$; each simulation follow the evolution of $256^3$, $512^3$ or $1024^3$ particles in a box of $1~h^{-1}{\rm Gpc}$ length. Billions of dark matter halos and cosmic voids have been identified in the simulations, whose runs required more than 35 million core hours. The \textsc{Quijote} simulations have been designed for two main purposes: 1) to quantify the information content on cosmological observables, and 2) to provide enough data to train machine learning algorithms. In this paper we describe the simulations and show a few of their applications. We also release the Petabyte of data generated, comprising hundreds of thousands of simulation snapshots at multiple redshifts, halo and void catalogs, together with millions of summary statistics such as power spectra, bispectra, correlation functions, marked power spectra, and estimated probability density functions.
\end{abstract}


\section{Introduction}
\label{sec:introduction}

The discovery of the accelerated expansion of the Universe \citep{Riess_1998, Perlmutter_1999} has revolutionized cosmology. We now believe that $\simeq70\%$ of the energy content of the Universe is made up of a mysterious substance that is accelerating the expansion of the Universe: dark energy. One of the most important tasks in modern cosmology is to unveil the properties of dark energy. This will help us to understand its nature and improve our knowledge on fundamental physics.

The spatial distribution of matter in the Universe is sensitive to the nature of dark energy, but also to other fundamental quantities such as the properties of dark matter, the sum of neutrino masses, and the initial conditions of the Universe. Thus, one of the most powerful ways to learn about fundamental physics, is by extracting that information from the large-scale structure of the Universe. This is the goal of many upcoming cosmological missions such as DESI\footnote{https://www.desi.lbl.gov}, Euclid\footnote{https://www.euclid-ec.org}, LSST\footnote{https://www.lsst.org}, PFS\footnote{https://pfs.ipmu.jp/index.html}, SKA\footnote{https://www.skatelescope.org}, and WFIRST\footnote{https://wfirst.gsfc.nasa.gov/index.html}.

The traditional way to retrieve information from cosmological observations is to compare summary statistics from data against theory predictions. An important question is then: What statistic or statistics should be used to extract the maximum information\footnote{By information we mean the constraints on the value of the cosmological parameters.} from cosmic observations?

It is well known that a Gaussian density field can be fully described by its power spectrum or correlation function \citep[see e.g.][]{Licia_2007, Ben_2013,Leclercq_2014}. This is the main reason why the power spectrum/correlation function is the most prominent statistic employed when analyzing cosmological data: at high-redshift, or on sufficiently large-scales at low-redshift, the Universe resembles a Gaussian density field, and most of the information embedded on it can be extracted from the power spectrum/correlation function.

The cosmic microwave background (CMB) is an example of a Gaussian density field\footnote{To-date, there is no significant evidence that points towards the CMB being non-Gaussian \citep{NonGaussinities_Planck}.}. All the information embedded in it can thus be retrieved through the power spectrum. Notice that for simplicity we are ignoring the non-Gaussian information that can be extracted from the CMB, e.g. through CMB lensing. Currently, some of the tightest and more robust constraints on the value of the cosmological parameters arise from CMB data \citep{Planck_2018}. Unfortunately, the primary CMB is limited to a plane on the sky at high-redshift, and is insensitive to low-redshift phenomena such as the transition from the matter dominated epoch to the Dark Energy dominated epoch.

Since the number of modes in 3-dimensional surveys is potentially much larger than in CMB observations, it is expected that the constraining power of those surveys will surpass those of CMB observations. Unfortunately, in 3-dimensional surveys, most of the modes are on mildly to non-linear scales. In the regime where the density field is non-Gaussian, it is currently unknown what is the statistic, or set of statistics, that will place the tightest constraints on the parameters. From a formal perspective, that question is also mathematically intractable. Being able to extract the cosmological information embedded into non-linear modes will enable us to tighten the value of the cosmological parameters and therefore to improve our understanding of fundamental physics.

One way to tackle this problem is to consider a given statistic/statistics and quantify the information content on it, from linear to non-linear scales. Numerical simulations are needed in this case, as they are one of the most powerful ways to obtain theory predictions in the fully non-linear regime, in real- and redshift-space, for any considered statistic. This is the motivation that brought us to develop the \textsc{Quijote} simulations; we designed them to allow the community to easily quantify the information content on different statistics into the fully non-linear regime.

Another way to approach the problem is to use advanced statistical techniques, such as machine/deep learning, to identify new and optimal statistics to extract cosmological information \citep{Siamak_2017, Charnock_2018,Justin_2019}. One of the requirements of these methods is to have a sufficiently large dataset to train the algorithms. The \textsc{Quijote} simulations have been designed to provide the community with a very big dataset of cosmological simulations.

In this paper we present the \textsc{Quijote} suite; the largest set of full N-body simulations\footnote{To the best of our knowledge.} run at this mass and spatial resolution to-date. The \textsc{Quijote} simulations contain 44100 full N-body simulations, expanding more than 7000 cosmological models and at a single redshift, contain more than 8.5 trillion particles. The computational cost of the simulations exceeds 35 million CPU hours, and over 1 Petabyte of data was generated.

We note that our simulations have a relatively low-resolution: they resolve halos with masses above $\simeq3\times10^{12}~h^{-1}M_\odot$ (high resolution) or $\simeq2\times10^{13}~h^{-1}M_\odot$ (fiducial resolution). Running the Quijote simulation suite at higher resolution would have been impossible due to computational and storage constraints.

The reason why we chose to run simulations at this resolution is twofold: 1) we need to run a large set of simulations to evaluate the Fisher matrix with all its ingredients converged, and 2) we wanted to sample a very large cosmological volume first, and then use machine learning to increase the resolution of the simulations (see below).

The Quijote simulations represent therefore a very useful tool to quantify information content on the matter field and for the most massive halos / most luminous galaxies. While the matter field is not directly observable in 3D (its projection is observed through weak lensing), quantifying the information content on different observables of it is still a useful exercise. If an observable provides competitive constraints on a given parameter for the matter field, it is worth exploring how much information will remain when considering galaxies. On the other hand, if a statistics does not accurately constraint the value of the parameters for the matter field, it is unlikely that it will do for galaxies.

This paper is organized as follows. In Section \ref{sec:Simulations} we describe in detail the \textsc{Quijote} simulations. We outline the data products generated by the simulations in Section \ref{sec:data_products}. We present a few applications of the \textsc{Quijote} simulations in Section \ref{sec:Applications}. In Section \ref{sec:resolution} we show several convergence tests in order to quantify the limitations of the simulations. Finally, we draw our conclusions in Section \ref{sec:Conclusions}.

\section{Simulations}
\label{sec:Simulations}

All the simulations in the \textsc{Quijote} suite are N-body simulations. They have been run using the TreePM code \textsc{Gadget-III}, an improved version of \textsc{Gadget-II} \citep{Gadget}.

The initial conditions of all simulations are generated at $z=127$. We obtain the input matter power spectrum and transfer functions by rescaling the $z=0$ matter power spectrum and transfer functions from CAMB \citep{CAMB}. For models with massive neutrinos we use the rescaling method developed in \cite{Zennaro_2017}, while for models with massless neutrinos we employ the traditional scale-independent rescaling
\begin{equation}
P_{\rm m}(k,z_i) = \left(\frac{D(z_i)}{D(z)}\right)^2P_{\rm m}(k,z=0)   , \hspace{0.5cm} f(z_i)\simeq\Omega_{\rm m}^\gamma(z_i)~,
\end{equation}
where $D(z)$ is the growth factor at redshift $z$, $f$ is the growth rate and $\gamma\simeq0.6$ for $\Lambda$CDM.
From the input matter power spectrum and transfer functions we compute displacements and peculiar velocities employing the Zeldovich approximation \citep{Zeldovich} (for cosmologies with massive neutrinos) or second order perturbation theory (for cosmologies with massless neutrinos). The displacements and peculiar velocities are then assigned to particles that are initially laid on a regular grid. In models with massive neutrinos we use two different grids that are offset by half a grid size: one grid for CDM and one grid for neutrinos. For 2LPT, we use the code in \url{https://cosmo.nyu.edu/roman/2LPT/}, while for neutrinos we used a modified version of N-GenIC, publicly available at \url{https://github.com/franciscovillaescusa/N-GenIC_growth}. The rescaling code used for massive neutrino cosmologies is publicly available in \url{https://github.com/matteozennaro/reps}.

All simulations have a cosmological volume of $1~(h^{-1}{\rm Gpc})^3$. The majority of the simulations follow the evolution of $512^3$ CDM particles (plus $512^3$ for simulations with massive neutrinos): our \textit{fiducial-resolution}. We however also have simulations with $256^3$ (\textit{low-resolution}) and $1024^3$ (\textit{high-resolution}) CDM particles. The gravitational softening length is set to $1/40$ of the mean interparticle distance, i.e. 100, 50 and 25 $h^{-1}{\rm kpc}$ for the low-, fiducial- and high-resolution simulations, respectively. The gravitational softening is the same for CDM and neutrino particles. We save snapshots at redshifts 0, 0.5, 1, 2, and 3. We also save the initial conditions and the scripts to generate them.

Table \ref{table:sims} summarizes the main features of all the \textsc{Quijote} simulations.

\begin{table*}
\begin{center}
\renewcommand{\arraystretch}{0.6}
\resizebox{1.0\textwidth}{!}{\begin{tabular}{| c || c | c | c | c | c | c | c | c || c | c | c | c | c |}
\hline
\multirow{2}{*}{Name} & \multirow{2}{*}{$\Omega_{\rm m}$} & \multirow{2}{*}{$\Omega_{\rm b}$} & \multirow{2}{*}{$h$} & \multirow{2}{*}{$n_{\rm s}$} & \multirow{2}{*}{$\sigma_8$} & \multirow{2}{*}{$M_\nu$(eV)} & \multirow{2}{*}{$w$} & \multirow{2}{*}{$\delta_{\rm b}$} & \multirow{2}{*}{realizations} & \multirow{2}{*}{simulations} & \multirow{2}{*}{ICs} & \multirow{2}{*}{$N_c^{1/3}$} & \multirow{2}{*}{$N_\nu^{1/3}$}\\
&&&&&&&&&&&&&\\
\hline
\hline
\multirow{7}{*}{Fid} & \multirow{7}{*}{\underline{0.3175}} & \multirow{7}{*}{\underline{0.049}} & \multirow{7}{*}{\underline{0.6711}} & \multirow{7}{*}{\underline{0.9624}} & \multirow{7}{*}{\underline{0.834}} & \multirow{7}{*}{\underline{0}} & \multirow{7}{*}{\underline{-1}} & \multirow{7}{*}{\underline{0}} & 15000 & standard & 2LPT & 512 & 0\\
&&&&&&&& &500 & standard & Zeldovich & 512 & 0 \\
&&&&&&&& &500 & paired fixed & 2LPT & 512 & 0\\
&&&&&&&& &1000 & standard & 2LPT & 256 & 0\\
&&&&&&&& &100 & standard & 2LPT & 1024 & 0\\
\hline
\multirow{2}{*}{$\Omega_{\rm m}^+$} & \multirow{2}{*}{\underline{0.3275}} & \multirow{2}{*}{0.049} & \multirow{2}{*}{0.6711} & \multirow{2}{*}{0.9624} & \multirow{2}{*}{0.834} & \multirow{2}{*}{0} & \multirow{2}{*}{-1} & \multirow{2}{*}{0} & 500 & standard & \multirow{2}{*}{2LPT} & \multirow{2}{*}{512} & \multirow{2}{*}{0} \\
&&&&&&&&& 500 & paired fixed & & & \\
\hline
\multirow{2}{*}{$\Omega_{\rm m}^-$} & \multirow{2}{*}{\underline{0.3075}} & \multirow{2}{*}{0.049} & \multirow{2}{*}{0.6711} & \multirow{2}{*}{0.9624} & \multirow{2}{*}{0.834} & \multirow{2}{*}{0} & \multirow{2}{*}{-1} & \multirow{2}{*}{0} & 500 & standard & \multirow{2}{*}{2LPT} & \multirow{2}{*}{512} & \multirow{2}{*}{0}\\
&&&&&&&&& 500 & paired fixed & &&\\
\hline
\multirow{2}{*}{$\Omega_{\rm b}^{++}$} & \multirow{2}{*}{0.3175} & \multirow{2}{*}{\underline{0.051}} & \multirow{2}{*}{0.6711} & \multirow{2}{*}{0.9624} & \multirow{2}{*}{0.834} & \multirow{2}{*}{0} & \multirow{2}{*}{-1} & \multirow{2}{*}{0} & 500 & standard & \multirow{2}{*}{2LPT} & \multirow{2}{*}{512} & \multirow{2}{*}{0}\\
&&&&&&&&& 500 & paired fixed & &&\\
\hline
\multirow{2}{*}{$\Omega_{\rm b}^+$} & \multirow{2}{*}{0.3175} & \multirow{2}{*}{\underline{0.050}} & \multirow{2}{*}{0.6711} & \multirow{2}{*}{0.9624} & \multirow{2}{*}{0.834} & \multirow{2}{*}{0} & \multirow{2}{*}{-1} & \multirow{2}{*}{0} & \multirow{2}{*}{500} & \multirow{2}{*}{paired fixed} & \multirow{2}{*}{2LPT} & \multirow{2}{*}{512} & \multirow{2}{*}{0}\\
&&&&&&&& && & &&\\
\hline
\multirow{2}{*}{$\Omega_{\rm b}^-$} & \multirow{2}{*}{0.3175} & \multirow{2}{*}{\underline{0.048}} & \multirow{2}{*}{0.6711} & \multirow{2}{*}{0.9624} & \multirow{2}{*}{0.834} & \multirow{2}{*}{0} & \multirow{2}{*}{-1} & \multirow{2}{*}{0} & \multirow{2}{*}{500} & \multirow{2}{*}{paired fixed} & \multirow{2}{*}{2LPT} & \multirow{2}{*}{512} & \multirow{2}{*}{0}\\
&&&&&&&& && & &&\\
\hline
\multirow{2}{*}{$\Omega_{\rm b}^{--}$} & \multirow{2}{*}{0.3175} & \multirow{2}{*}{\underline{0.047}} & \multirow{2}{*}{0.6711} & \multirow{2}{*}{0.9624} & \multirow{2}{*}{0.834} & \multirow{2}{*}{0} & \multirow{2}{*}{-1} & \multirow{2}{*}{0} & 500 & standard & \multirow{2}{*}{2LPT} & \multirow{2}{*}{512} & \multirow{2}{*}{0}\\
&&&&&&&&& 500 & paired fixed & &&\\
\hline
\multirow{2}{*}{$h^+$} & \multirow{2}{*}{0.3175} & \multirow{2}{*}{0.049} & \multirow{2}{*}{\underline{0.6911}} & \multirow{2}{*}{0.9624} & \multirow{2}{*}{0.834} & \multirow{2}{*}{0} & \multirow{2}{*}{-1} & \multirow{2}{*}{0} & 500 & standard & \multirow{2}{*}{2LPT} & \multirow{2}{*}{512} & \multirow{2}{*}{0}\\
&&&&&&&&& 500 & paired fixed & &&\\
\hline
\multirow{2}{*}{$h^-$} & \multirow{2}{*}{0.3175} & \multirow{2}{*}{0.049} & \multirow{2}{*}{\underline{0.6511}} & \multirow{2}{*}{0.9624} & \multirow{2}{*}{0.834} & \multirow{2}{*}{0} & \multirow{2}{*}{-1} & \multirow{2}{*}{0} & 500 & standard & \multirow{2}{*}{2LPT} & \multirow{2}{*}{512} & \multirow{2}{*}{0}\\
&&&&&&&&& 500 & paired fixed & &&\\
\hline
\multirow{2}{*}{$n_{\rm s}^+$} & \multirow{2}{*}{0.3175} & \multirow{2}{*}{0.049} & \multirow{2}{*}{0.6711} & \multirow{2}{*}{\underline{0.9824}} & \multirow{2}{*}{0.834} & \multirow{2}{*}{0} & \multirow{2}{*}{-1} & \multirow{2}{*}{0} & 500 & standard & \multirow{2}{*}{2LPT} & \multirow{2}{*}{512} & \multirow{2}{*}{0}\\
&&&&&&&&& 500 & paired fixed & &&\\
\hline
\multirow{2}{*}{$n_{\rm s}^+$} & \multirow{2}{*}{0.3175} & \multirow{2}{*}{0.049} & \multirow{2}{*}{0.6711} & \multirow{2}{*}{\underline{0.9424}} & \multirow{2}{*}{0.834} & \multirow{2}{*}{0} & \multirow{2}{*}{-1} & \multirow{2}{*}{0} & 500 & standard & \multirow{2}{*}{2LPT} & \multirow{2}{*}{512} & \multirow{2}{*}{0}\\
&&&&&&&&& 500 & paired fixed & & & \\
\hline
\multirow{2}{*}{$\sigma_8^+$} & \multirow{2}{*}{0.3175} & \multirow{2}{*}{0.049} & \multirow{2}{*}{0.6711} & \multirow{2}{*}{0.9624} & \multirow{2}{*}{\underline{0.849}} & \multirow{2}{*}{0} & \multirow{2}{*}{-1} & \multirow{2}{*}{0} & 500 & standard & \multirow{2}{*}{2LPT} & \multirow{2}{*}{512} & \multirow{2}{*}{0}\\
&&&&&&&&& 500 & paired fixed & &&\\
\hline
\multirow{2}{*}{$\sigma_8^-$} & \multirow{2}{*}{0.3175} & \multirow{2}{*}{0.049} & \multirow{2}{*}{0.6711} & \multirow{2}{*}{0.9624} & \multirow{2}{*}{\underline{0.819}} & \multirow{2}{*}{0} & \multirow{2}{*}{-1} & \multirow{2}{*}{0} & 500 & standard & \multirow{2}{*}{2LPT} & \multirow{2}{*}{512} & \multirow{2}{*}{0}\\
&&&&&&&&& 500 & paired fixed & &&\\
\hline
\multirow{2}{*}{$M_\nu^{+++}$} & \multirow{2}{*}{0.3175} & \multirow{2}{*}{0.049} & \multirow{2}{*}{0.6711} & \multirow{2}{*}{0.9624} & \multirow{2}{*}{0.834} & \multirow{2}{*}{\underline{0.4}} & \multirow{2}{*}{-1} & \multirow{2}{*}{0} & 500 & standard & \multirow{2}{*}{Zeldovich} & \multirow{2}{*}{512} & \multirow{2}{*}{512}\\
&&&&&&&&& 500 & paired fixed & &&\\
\hline
\multirow{2}{*}{$M_\nu^{++}$} & \multirow{2}{*}{0.3175} & \multirow{2}{*}{0.049} & \multirow{2}{*}{0.6711} & \multirow{2}{*}{0.9624} & \multirow{2}{*}{0.834} & \multirow{2}{*}{\underline{0.2}} & \multirow{2}{*}{-1} & \multirow{2}{*}{0} & 500 & standard & \multirow{2}{*}{Zeldovich} & \multirow{2}{*}{512} & \multirow{2}{*}{512}\\
&&&&&&&&& 500 & paired fixed & &&\\
\hline
\multirow{2}{*}{$M_\nu^+$} & \multirow{2}{*}{0.3175} & \multirow{2}{*}{0.049} & \multirow{2}{*}{0.6711} & \multirow{2}{*}{0.9624} & \multirow{2}{*}{0.834} & \multirow{2}{*}{\underline{0.1}} & \multirow{2}{*}{-1} & \multirow{2}{*}{0} & 500 & standard & \multirow{2}{*}{Zeldovich} & \multirow{2}{*}{512} & \multirow{2}{*}{512}\\
&&&&&&&&& 500 & paired fixed & &&\\
\hline
\multirow{2}{*}{$w^+$} & \multirow{2}{*}{0.3175} & \multirow{2}{*}{0.049} & \multirow{2}{*}{0.6711} & \multirow{2}{*}{0.9624} & \multirow{2}{*}{0.834} & \multirow{2}{*}{0} & \multirow{2}{*}{\underline{-1.05}} & \multirow{2}{*}{0} & \multirow{2}{*}{500} & \multirow{2}{*}{standard} & \multirow{2}{*}{Zeldovich} & \multirow{2}{*}{512} & \multirow{2}{*}{0}\\
&&&&&&&&& & & &&\\
\hline
\multirow{2}{*}{$w^-$} & \multirow{2}{*}{0.3175} & \multirow{2}{*}{0.049} & \multirow{2}{*}{0.6711} & \multirow{2}{*}{0.9624} & \multirow{2}{*}{0.834} & \multirow{2}{*}{0} & \multirow{2}{*}{\underline{-0.95}} & \multirow{2}{*}{0} & \multirow{2}{*}{500} & \multirow{2}{*}{standard} & \multirow{2}{*}{Zeldovich} & \multirow{2}{*}{512} & \multirow{2}{*}{0}\\
&&&&&&&&&&&&&\\
\hline
\multirow{2}{*}{${\rm DC}^+$} & \multirow{2}{*}{0.3175} & \multirow{2}{*}{0.049} & \multirow{2}{*}{0.6711} & \multirow{2}{*}{0.9624} & \multirow{2}{*}{0.834} & \multirow{2}{*}{0} & \multirow{2}{*}{-1} & \multirow{2}{*}{\underline{0.035}} & \multirow{2}{*}{500} & \multirow{2}{*}{standard} & \multirow{2}{*}{2LPT} & \multirow{2}{*}{512} & \multirow{2}{*}{0}\\
&&&&&&&&&&&&&\\
\hline
\multirow{2}{*}{${\rm DC}^-$} & \multirow{2}{*}{0.3175} & \multirow{2}{*}{0.049} & \multirow{2}{*}{0.6711} & \multirow{2}{*}{0.9624} & \multirow{2}{*}{0.834} & \multirow{2}{*}{0} & \multirow{2}{*}{-1} & \multirow{2}{*}{\underline{-0.035}} & \multirow{2}{*}{500} & \multirow{2}{*}{standard} & \multirow{2}{*}{2LPT} & \multirow{2}{*}{512} & \multirow{2}{*}{0}\\
&&&&&&&&&&&&&\\
\hline
\hline
\multirow{3}{*}{LH} & \multirow{3}{*}{[0.1 , 0.5]} & \multirow{3}{*}{[0.03 , 0.07]} & \multirow{3}{*}{[0.5 , 0.9]} & \multirow{3}{*}{[0.8 , 1.2]} & \multirow{3}{*}{[0.6 , 1.0]} & \multirow{3}{*}{0} & \multirow{3}{*}{-1} & \multirow{2}{*}{0} & 2000 & standard & \multirow{3}{*}{2LPT} & 512 & \multirow{3}{*}{0}\\
&&&&&&&&& 2000 & fixed & &512&\\
&&&&&&&&& 2000 & standard & &1024&\\
\hline
\multirow{2}{*}{LH$\nu w$} & \multirow{2}{*}{[0.1 , 0.5]} & \multirow{2}{*}{[0.03 , 0.07]} & \multirow{2}{*}{[0.5 , 0.9]} & \multirow{2}{*}{[0.8 , 1.2]} & \multirow{2}{*}{[0.6 , 1.0]} & \multirow{2}{*}{[0 , 1]} & \multirow{2}{*}{[-1.3 , -0.7]} & \multirow{2}{*}{0} & \multirow{2}{*}{5000} & \multirow{2}{*}{standard} & \multirow{2}{*}{Zeldovich} & \multirow{2}{*}{512} & \multirow{2}{*}{512}\\
&&&&&&&&& &  & &&\\

\hline
\hline
\multirow{4}{*}{total} & \multirow{2}{*}{-} & \multirow{2}{*}{-} & \multirow{2}{*}{-} & \multirow{2}{*}{-} & \multirow{2}{*}{-} & \multirow{2}{*}{-} & \multirow{2}{*}{-} & \multirow{2}{*}{-} &  \multirow{2}{*}{44100} & \multirow{2}{*}{-} & \multirow{2}{*}{-} & \multirow{2}{*}{-} & \multirow{2}{*}{-}\\
&&&&&&&&&&&&&\\
\cline{2-14}
 & \multirow{2}{*}{-} & \multirow{2}{*}{-} & \multirow{2}{*}{-} & \multirow{2}{*}{-} & \multirow{2}{*}{-} & \multirow{2}{*}{-} & \multirow{2}{*}{-} &  \multirow{2}{*}{-} & \multirow{2}{*}{-} & \multirow{2}{*}{-} & \multirow{2}{*}{-} & \multirow{2}{*}{19811} & \multirow{2}{*}{10240}\\
&&&&&&&&&&&&&\\
\hline
\end{tabular}}
\end{center}
\caption{Characteristics of the \textsc{Quijote} simulations. The simulations in the first block have been designed to quantify the information content on cosmological observables. The have a large number of realizations for a fiducial cosmology (needed to estimate the covariance matrix) and simulations varying just one cosmological parameter at a time (needed to compute numerical derivatives). The simulations in the second block arise from latin-hypercubes expanding a large volume in parameter space. The initial conditions of all simulations were generated at $z=127$ using 2LPT, except for the simulations with massive neutrinos, where we used the Zel'dovich approximation. All simulations have a volume of $1~(h^{-1}{\rm Gpc})^3$ and we have three different resolutions: low-resolution ($256^3$ particles), fiducial-resolution ($512^3$ particles) and high-resolution ($1024^3$ particles). In the simulations with massive neutrinos we assume three degenerate neutrino masses. Simulations have been run with the TreePM+SPH  {\sc Gadget-III}  code. We save snapshots at redshifts 3, 2, 1, 0.5 and 0. The parameter $\delta_b$ stands for background overdensity, and it only applies to the separate Universe simulations.}
\label{table:sims}
\end{table*}

\subsection{Simulations with massive neutrinos}

In simulations with massive neutrinos, we use the traditional particle-based method \citep{Brandbyge_2008, Viel_2010} to model the cosmic neutrino background. In that method, neutrinos are described as a collisionless and pressureless fluid, that is discretized into a set of neutrino particles. Those particles are assigned thermal velocities (on top of peculiar velocities) that are randomly drawn from their Fermi-Dirac distribution at the simulation starting redshift.

One of the well-known problems of this method, is that a significant fraction of the neutrino particles will cross the simulation box several times (due to their large thermal velocities). This will erase the clustering of neutrinos on small scales, producing a white power spectrum (or shot-noise). This effect is however negligible on most of the observational quantities, e.g. the total matter power spectrum, the halo/galaxy power spectrum.

New methods have been developed to address this problem \citep[see e.g.][]{Arka_18}. The 5000 simulations of the LH$\nu w$ latin-hypercube have been run using this method, that provides a neutrino density field with a negligible level of shot-noise.

\subsection{Paired fixed simulations}

The \textsc{Quijote} simulations contain a) standard, b) fixed and c) paired fixed simulations. The differences between those is the way the initial conditions are generated. Consider a Fourier-space mode, $\delta(\vec{k})$. Since it is in general a complex number, we can write it as $\delta(\vec{k})=Ae^{i\theta}$, where both the amplitude $A$, and the phase $\theta$, depends on the considered wavenumber $\vec{k}$. For Gaussian density fields, $A$ follows a Rayleigh distribution and $\theta$ is drawn from an uniform distribution between 0 and $2\pi$. This is the standard way to generate initial conditions for cosmological simulations. In fixed simulations, while $\theta$ is still drawn from a uniform distribution between 0 and $2\pi$, the value of $A$ is fixed to the square root of the variance of the previous Rayleigh distribution. Finally, paired fixed simulations are two fixed simulations where the phases of the two pairs differ by $\pi$. We refer the reader to \cite{Pontzen_2016,RP_16,Paco_2018b} for further details.

Fixed and paired fixed simulations have received a lot of attention recently, since it has been shown that they can significantly reduce the amplitude of cosmic variance on different statistics (e.g. the power spectrum) without
inducing a bias on the results \citep{Pontzen_2016,RP_16,Paco_2018b,Lauren_19,Albert_19,Klypin_19}. While these simulations can not be used to estimate covariance matrices, they may be useful to compute numerical derivatives, or to provide an effective larger cosmological volume. For this reason, some of the simulations we have run are fixed and paired fixed.

\subsection{Separate Universe simulations \& Super-sample effect}

In a conventional simulation, it is assumed that the mean density in the
box is equal to that of the whole observable universe.
So all the above simulations have a mean background overdensity,
$\delta_b$, equal to 0.
This effectively truncates the sampling of matter power spectrum at the
box scale.
In reality, it is expected that regions of $1~(h^{-1}{\rm
Gpc})^3$, as the ones simulated by the Quijote suite, will have
background densities different to 0, due to the long-wavelength modes
beyond the box size.

The non-vanishing $\delta_b$ modulates the local growth factor and
expansion rate, known as the growth and dilation effects \citep{Li_2014,
Li_2018}, respectively.
As an example, a positive $\delta_b$ enhances the growth of structures,
and at the same time slows down the expansion of the local region as
compared to the global universe.
Together the growth and dilation effects are known as the super-sample
effect, given their origin from the long-wavelength modes beyond the
sampled (simulation or survey) region.
To be able to quantify the impact of the super-sample effect, we have
run a set of 1000 simulations with non-vanishing $\delta_b$, using a
technique called the separate Universe simulation.

In a $\Lambda$CDM universe, there is a well-known symmetry that allows
one to absorb the mean overdensity into a redefinition of cosmology with
curvature \citep{Sirko05}.
This allows us to use existing code to perform the simulations while
only modifying their cosmological parameters.
In particular we care about the linear response to $\delta_b$, which we
can measure with central difference of a pair of separate universe
simulations.
We have run two sets of 500 simulations each, with the fiducial
cosmology and $\delta_b=\pm0.035$, and refer the readers to
\citet{Li_2014} for more details on the setup of the paired simulations.

An import consequence of the super-sample effect is that it introduces
additional covariance of generic observables, called the super-sample
covariance \citep{TakadaHu_2013, Li_2018, Philcox_2020}.
This extra covariance arises from the coherent modulation of the long
modes on all observables within the local region, and the unknown nature
of $\delta_b$.
Intuitively, the super-sample covariance between two observables $O_1$
and $O_2$ is
\begin{equation}
  C^\textsc{ssc}_{\scriptstyle o_1 o_2}
  = \sigma_b^2 \frac{\mathrm{d} O_1}{\mathrm{d} \delta_b}
               \frac{\mathrm{d} O_2}{\mathrm{d} \delta_b},
\end{equation}
where $\sigma_b^2$ is the variance of $\delta_b$, and the other two
terms are the linear response of the observables.
This allows us to evaluate the impact of the super-sample covariance
using the Fisher matrix formalism.

\subsection{Fiducial cosmology}

The value of the cosmological parameters for our fiducial model are: $\Omega_{\rm m}=0.3175$, $\Omega_{\rm b}=0.049$, $h=0.6711$, $n_s=0.9624$, $\sigma_8=0.834$, $M_\nu=0.0$ eV, and $w=-1$. The values of those parameters are in good agreement with the latest constraints by Planck \citep{Planck_2018}.

For this model, we have run a total 17100 simulations. Of those, 15000 are standard simulations run at the fiducial resolution with 2LPT initial conditions. The main purpose of these simulations is to compute covariance matrices. We also have a set of 500 paired fixed simulations, with 2LPT initial conditions at fiducial resolution that can be used to study properties of paired fixed simulations and to compute numerical derivatives.

Furthermore, we have a set of 500 standard simulations with Zel'dovich initial conditions at fiducial resolution needed to compute the derivatives with respect to neutrino masses (see subsection \ref{subsec:ZA_vs_2LPT}). Finally, a set of 1000 standard simulations at low-resolution, and 100 standard simulations at high-resolution are available to carry out resolution tests and apply super-resolution techniques (see subsection \ref{subsec:superresolution}).

\subsection{Simulations for numerical derivatives}

One of the ingredients needed to quantify the information content of a statistic is the partial derivatives of it with respect to the cosmological parameters (see subsection \ref{subsec:Information_content}). For $\Omega_{\rm m}$, $\Omega_{\rm b}$, $h$, $n_s$, $\sigma_8$, and $w$ we compute partial derivatives as
\begin{equation}
\frac{\partial \vec{S}}{\partial \theta}\simeq\frac{\vec{S}(\theta+d\theta)-\vec{S}(\theta-d\theta)}{2d\theta}~,
\label{eq:finite_differences}
\end{equation}
where $\vec{S}$ is the considered statistic (e.g. the matter power spectrum at different wavenumbers), and $\theta$ is the cosmological parameter. We thus need to evaluate the statistic on simulations where only the considered parameter is varied above and below its fiducial value. In order to fulfill this requirement, we have run simulations varying only one cosmological parameter at a time. For instance, the simulations coined $\Omega_{\rm m}^+$/$\Omega_{\rm m}^-$ have the same value $\Omega_{\rm b}$, $h$, $n_s$, $\sigma_8$, $M_\nu$ and $w$ as the fiducial model but the value of $\Omega_{\rm m}$ is slightly larger/smaller. In this case $d\Omega_{\rm m}/\Omega_{\rm m}\simeq1.8\%$: $\Omega_{\rm m}=0.3275$ for $\Omega_{\rm m}^+$ and $\Omega_{\rm m}=0.3075$ for $\Omega_{\rm m}^-$.

In the simulations $\Omega_{\rm b}^{++}$ and $\Omega_{\rm b}^{--}$ we vary $\Omega_{\rm b}$ by $d\Omega_{\rm b}/\Omega_{\rm b}\simeq4\%$. While when varying the others parameters $h$, $n_s$, $\sigma_8$ and $w$ we have $dh/h\simeq3\%$, $dn_s/n_s\simeq2\%$, $d\sigma_8/\sigma_8\simeq1.8\%$, $dw/w=5\%$, respectively. These numbers were chosen such as the difference is small enough to approximate the derivatives, but not too small to be dominated by numerical noise. In the $\Omega_{\rm b}^{+}$ and $\Omega_{\rm b}^{-}$ simulations we have $d\Omega_{\rm b}/\Omega_{\rm b}\simeq2\%$. For most of the statistics we have considered, this difference is too small and the derivatives are slightly affected by numerical noise.

For all these models, we have run 500 standard simulations and 500 paired fixed simulations using 2LPT at the fiducial resolution. The exception is the models with $w\neq-1$ where we only run 500 standard simulations and $\Omega_{\rm b}^{+}$/$\Omega_{\rm b}^{-}$ that only have 500 paired fixed simulations.

\begin{figure*}
\begin{center}
\includegraphics[width=1.0\textwidth]{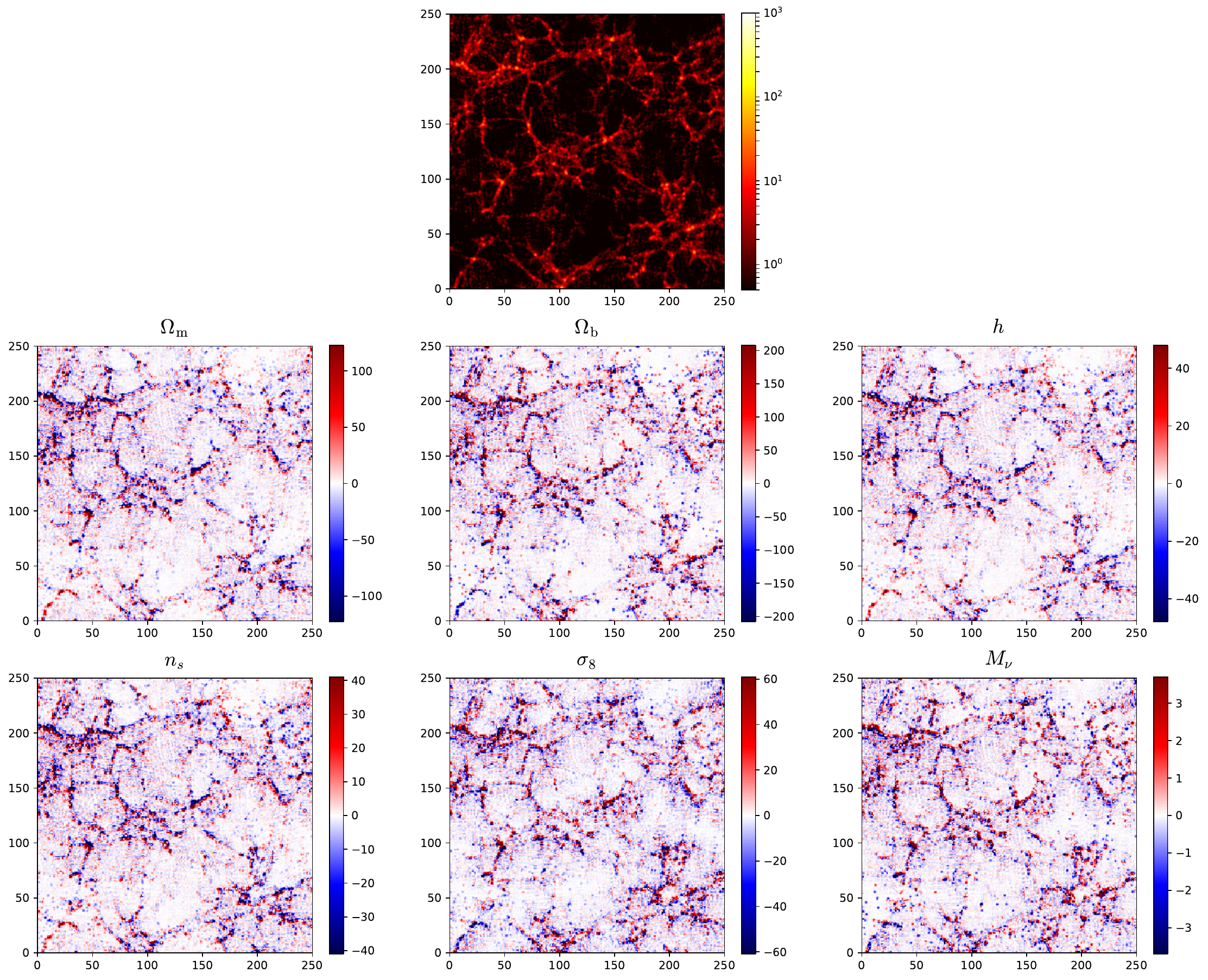}
\caption{The image on the top shows the large-scale structure in a region of $250\times250\times15~(h^{-1}{\rm Mpc})^3$ at $z=0$ for the fiducial cosmology. We have taken simulations with the same random seed but different values of just one single parameter, and used them to compute the derivative of the density field with respect to the parameters. The panels on the middle and bottom row show those derivatives with respect to $\Omega_{\rm m}$ (middle-left), $\Omega_{\rm b}$ (middle-center), $h$ (middle-right), $n_s$ (bottom-left), $\sigma_8$ (bottom-middle), and $M_\nu$ (bottom-right). It can be seen how different parameters affect the large-scale structure in different manners. For instance, the filament on the bottom-left part of the plot responses differently to each parameter. Neural networks can use these features to extract information from non standard summary statistics.}
\label{fig:Derivative_image}
\end{center}
\end{figure*}

To compute numerical derivatives with respect to massive neutrinos, we cannot use Eq. \ref{eq:finite_differences}, since the second term in the numerator will correspond to a Universe with negative neutrino masses\footnote{Notice that our fiducial cosmology is for a Universe with massless neutrinos.}. For this reason, we have run simulations at several values of the neutrino masses: $M_\nu^{+}=0.1$ eV, $M_\nu^{++}=0.2$ eV and $M_\nu^{+++}=0.4$ eV. From these simulations, several derivatives can be computed
\begin{eqnarray}
\frac{\partial \vec{S}}{\partial M_\nu}&\simeq&\frac{\vec{S}(M_\nu)-\vec{S}(M_\nu=0)}{M_\nu}\nonumber\\
\frac{\partial \vec{S}}{\partial M_\nu}&\simeq&\frac{-\vec{S}(2M_\nu) + 4\vec{S}(M_\nu) - 3\vec{S}(M_\nu=0)}{2M_\nu}\nonumber\\
\frac{\partial \vec{S}}{\partial M_\nu}&\simeq&\frac{\vec{S}(4M_\nu) - 12\vec{S}(2M_\nu) + 32\vec{S}(M_\nu) - 21\vec{S}(M_\nu=0)}{12M_\nu}\nonumber
\end{eqnarray}
where the first equation can be used for $M_\nu=0.1$, 0.2 or 0.4 eV. The second equation can instead be evaluated with $M_\nu=0.1$ or 0.2 eV while the last equation requires $M_\nu=0.1$ eV. Notice that if the differences between the fiducial model and the cosmology with 0.1 eV neutrinos are not dominated by noise, the last equation will provide the most precise estimation of the derivative. In some cases, e.g. with the halo mass function, differences between the fiducial model with massless neutrinos and cosmology with 0.1 eV neutrinos is too small, and therefore dominated by noise. In these cases, it is recommended to use the above second equation with $M_\nu=0.2$ eV.

For the models with 0.1 eV, 0.2 eV and 0.4 eV we have run 500 standard and 500 paired fixed simulations at the fiducial resolution. As stated above, for models with massive neutrinos, the initial conditions have been generated using the Zel'dovich approximation.

The top panel of Fig. \ref{fig:Derivative_image} shows the spatial distribution of matter in a realization of the fiducial cosmology. The other panels show the derivative of the density field of that particular realization with respect to the parameters $\Omega_{\rm m}$, $\Omega_{\rm b}$, $h$, $n_s$, $\sigma_8$, and $M_\nu$. It can be seen how the morphology of the density field responses differently to each cosmological parameters. Neural networks are trained to identify these changes and constrain parameter values using them.

\subsection{Latin-hypercubes}
\label{subsec:LH}

Besides the simulations described above, we have also run a set of 11000 simulations on different latin-hypercubes. The main purpose of these simulations is to provide enough data to train machine learning algorithms. In Section \ref{sec:Applications} we outline some applications of these simulations.

The simulations can be split into two main sets. In the first one, called LH, we use a latin-hypercube where we vary the value of $\Omega_{\rm m}$ between 0.1 and 0.5, $\Omega_{\rm b}$ between 0.03 and 0.07, $h$ between 0.5 and 0.9, $n_s$ between 0.8 and 1.2, $\sigma_8$ between 0.6 and 1.0 and keep fixed $M_\nu$ to 0.0 eV and $w$ to -1. LH is made of three different sets, but the value of the cosmological parameters is the same among the three set. The first one, is made of standard simulations with different random seeds. The second one, is made of fixed simulations, all of them having the same random seed. Those two sets have been run at the fiducial resolution. The last set is made of standard simulations with different random seeds but at high-resolution: $1024^3$ CDM particles. The initial conditions of all these simulations were generated with 2LPT. The three different sets contain 2000 simulations each. The set with fixed simulations can be used to create an accurate emulator, while the other two sets can be used to train machine learning algorithms accounting for the presence of cosmic variance.

\begin{figure*}
\begin{center}
\includegraphics[width=0.49\textwidth]{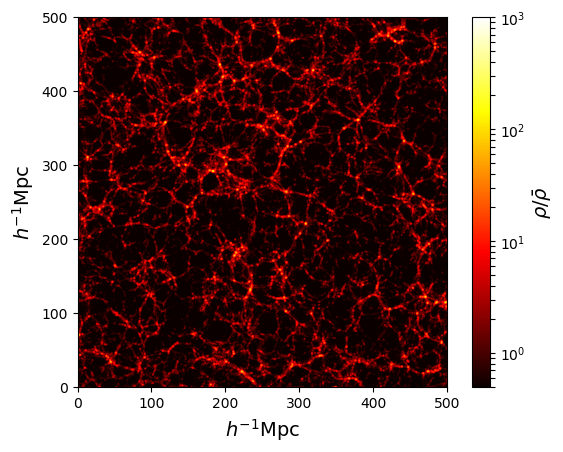}
\includegraphics[width=0.49\textwidth]{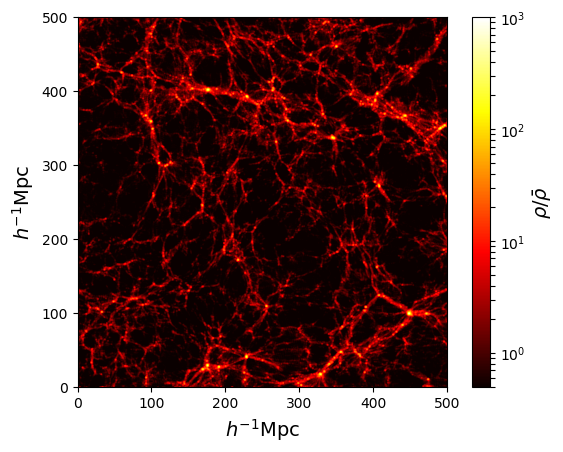}\\
\includegraphics[width=0.49\textwidth]{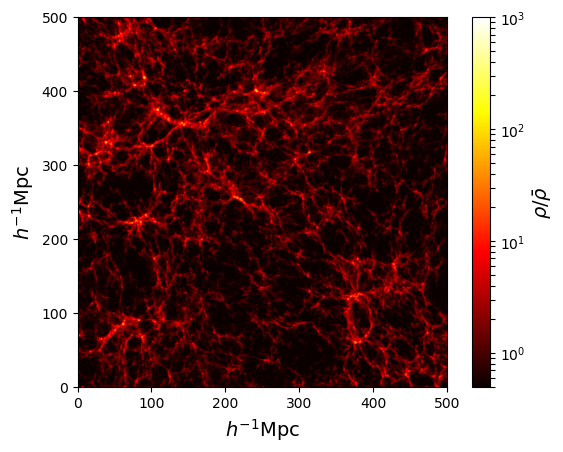}
\includegraphics[width=0.49\textwidth]{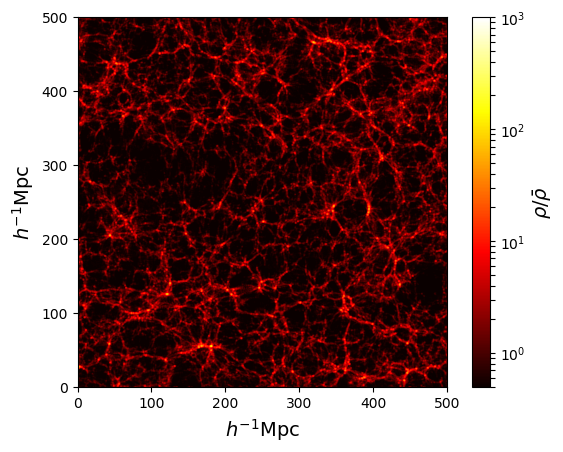}\\
\includegraphics[width=0.49\textwidth]{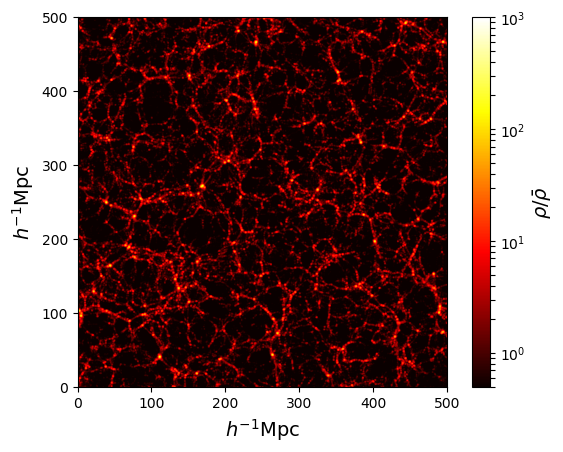}
\includegraphics[width=0.49\textwidth]{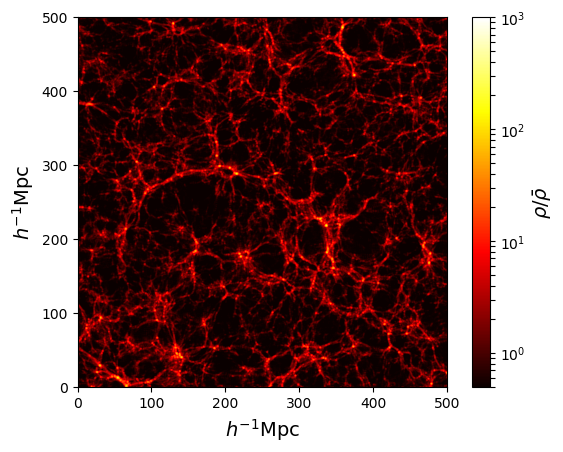}
\caption{Projected density field of a region of $500\times500\times10~(h^{-1}{\rm Mpc})^3$ from 6 different cosmologies of the high-resolution LH simultions at $z=0$. The top-left panel corresponds to a model close to Planck: $\{\Omega_{\rm m}, \Omega_{\rm b}, h, n_{\rm s}, \sigma_8\}$=\{0.3223, 0.04625, 0.7015, 0.9607, 0.8311\}. The other panels represent cosmologies with \{0.1005, 0.04189, 0.5133, 1.0107,  0.9421\} (top-right), \{0.1029, 0.06613, 0.7767, 1.0115, 0.6411\} (middle-left), \{0.1645, 0.05257, 0.7743, 1.0311, 0.6149\} (middle-right), \{0.4487, 0.03545, 0.5167, 1.0387, 0.9291\} (bottom-left), \{0.1867, 0.04503, 0.6189, 0.8307, 0.7187\} (bottom-right). These images show the wide range in parameter variations the Quijote simulations cover; some cosmologies exhibit very tick and/or long filaments, while others exhibit unusual clustering patterns.}
\label{fig:images}
\end{center}
\end{figure*}

The second latin-hypercube, called LH$\nu w$, is a set 5000 standard simulations with different random seeds where the value of the cosmological parameters is changed within: $\Omega_{\rm m}\in[0.1, 0.5]$, $\Omega_{\rm b}\in[0.03, 0.07]$, $h\in[0.5, 0.9]$, $n_s\in[0.8, 1.2]$, $\sigma_8\in[0.6, 1.0]$, $M_\nu\in[0, 1]$, $w\in[-1.3, -0.7]$. Since these simulations contain massive neutrinos, the initial conditions were generated using the Zel'dovich approximation. All the simulations follow the evolution of $512^3$ CDM particles plus $512^3$ neutrino particles. Each simulation is run with a different random seed.

Fig. \ref{fig:images} shows the spatial distribution of matter in 6 different cosmological models of the high-resolution LH simulations at $z=0$. Different features show up in the different images: from very long and thick filaments to highly clustered structures. This reflects the broad range covered by the \textsc{Quijote} simulations in the parameter space; from realistic models to extreme scenarios.

\section{Data products}
\label{sec:data_products}

In this section we describe the data products of the \textsc{Quijote} simulations.

\subsection{Snapshots}

We provide access to the full snapshots of the simulations at redshifts 0, 0.5, 1, 2, 3, and the initial conditions at $z=127$. The snapshots only have four different fields, 1) header, 2) positions, 3) velocities and 4) IDs.

The header contains information about the snapshot such as, redshift, value of $\Omega_{\rm m}$, $\Omega_\Lambda$, number of particles, number of files...etc. The position block stores the positions of the particles in comoving $h^{-1}{\rm kpc}$. The velocities of the particles are in the velocities block while the IDs block hosts the unique ids of the particles. The positions and velocities are saved as 32-bits floats, while the IDs are 32-bits integers. The snapshots are stored in either \textsc{Gadget-II} or hdf5 format. \textsc{Pylians}\footnote{\url{https://github.com/franciscovillaescusa/Pylians}} can be used to read the snapshots, independently of the format.

\subsection{Halo catalogues}

We save halo catalogues at each redshift for all the simulations; a total of 215500 halo catalogues. Halos are identified using the Friends-of-Friends (FoF) algorithm \citep{FoF}. We set the value of the linking length parameter to $b=0.2$. Each halo catalogue contains the positions, velocities, masses and total number of particles of each halo. Only CDM particles are linked in the FoF halos, as the contribution of neutrinos to the total halo mass is expected to be negligible \citep{Paco_11,Paco_12,Ichiki_Takada_2011, LoVerde_2014}. For simulations with large neutrino masses (e.g. the simulations in the LH$\nu w$) we also provide halo catalogues with the mass of halos being CDM plus neutrinos. The halo catalogues are saved in a binary format. \textsc{Pylians} can be used to read the catalogues. We only save halos that contain at least 20 CDM particles. We have not carried out an unbinding step to remove spurious halos. We thus caution the user to take this into account when using FoF halos with low number of particles.

For a subset of the simulations we have also identified halos using the \textsc{AMIGA} halo finder \citep{AMIGA}.

\subsection{Void catalogues}
\label{subsec:void_catalogues}

We provide void catalogs from every simulation at each redshift. For simulations with massive neutrinos, we provide two void catalogs - one in which the voids were identified using the total matter field, and the other in which the voids were identified in only the CDM+baryon field. For cosmologies with massless neutrinos, we only provide the latter. More than 250000 void catalogues are thus provided by the \textsc{Quijote} simulations.

Voids are identified in the simulations using the void finder used in \citet{Banerjee_2016}. The algorithm is as follows. First, the relevant overdensity field (CDM or CDM+neutrinos) is computed on a regular grid. This overdensity field is then smoothed on some scale $R_{\rm smooth}$ using a top-hat filter. All voxels at which the value of the smoothed overdensity field is below some threshold $\delta_{\rm threshold}$ are stored. Note that the initial $R_{\rm smooth}$ is chosen to be quite large ($\sim 100~h^{-1}{\rm Mpc}$). The grid voxels are then sorted in order of increasing overdensity, and the voxel with the lowest overdensity (or most underdense) is labelled as a void center with void radius $R_{\rm smooth}$. Since we use spherical top-hat smoothing, we can also associate a mass with the void: $M_{v} = 4/3\pi R_{\rm smooth}^3\bar \rho (1+\delta_{\rm threshold})$.
We also tag all voxels within radius $R_{\rm smooth}$ so that they cannot later be labelled as void centers. We then work down the list of points which crossed the threshold, i.e. to higher overdensities (less underdense), identifying them as new void centers with radius $R_{\rm smooth}$ if they do not overlap with previously identified voids.

Once all voxels below threshold for a given $R_{\rm smooth}$ have been checked, we move to a smaller value of $R_{\rm smooth}$ and repeat the procedure outlined above. In this way, the largest voids are the first identified, and then progressively smaller voids are found and stored in the void catalog. Note that by this definition, we do not have nested void regions in the provided void catalogs.

By default, our void find was run using $\delta_{\rm threshold}=-0.7$, but we also provide void catalogues with different values of $\delta_{\rm threshold}$ such as -0.5 or -0.3. Our void catalogs contain the positions, radii, and void size functions (number density of voids per unit of radius). The void catalogues are stored in HDF5 files.

\subsection{Power spectra}

We compute power spectra for 1) total matter field, 2) CDM+baryons (only for simulations with massive neutrinos), and 3) halos with different masses. The power spectra are computed for all simulations, at the redshifts 0, 0.5, 1, 2, and 3, and also at $z=127$ for 1) and 2). We compute the power spectra in both real- and redshift-space. In redshift-space, we place the redshift-space distortions along one cartesian axis, and compute the monopole, quadrupole and hexadecapole. We repeat the procedure for the three cartesian axes, i.e. in redshift-space, we compute three power spectra instead of one. In total, the \textsc{Quijote} simulations contain over 1 million power spectra.

\subsection{Marked power spectra}

Marked correlation functions are special 2-point statistics were correlations are weighted according to a mark, e.g. some environmental property. They have been shown to be interesting tools to study galaxy clustering dependence on galaxy properties such as morphology, luminosity, etc. \citep{Beisbart_2000,Sheth_2005}, halo clustering dependence on merger history \citep{Gottloeber_2002}, modified theories of gravity \citep{White_2016,Valogiannis_2017,Armijo_2018,Hernandez-Aguayo_2018}, and neutrinos' masses \citep{Massara_19}.

We compute marked power spectra of the matter density field, which are the Fourier counterpart of marked correlations. Inspired by~\cite{White_2016}, we consider the mark $M(\vec{x})$ of the form
\begin{equation}
    M(\vec{x}) = \left[ \frac{1+\delta_s}{1+\delta_s+\delta_R(\vec{x})}\right]^p\, ,
\end{equation}
that depends on the local matter density $\delta_R(\vec{x})$, a parameter $\delta_s$, and an exponent $p$. The density $\delta_R(\vec{x})$ is obtained by smoothing the matter density field with a Top-Hat filter at scale $R$ and can be evaluated at each point in the space $\vec{x}$. Thus, the mark depends on three parameters: $R$, $p$, and $\delta_s$. When $\delta_s \rightarrow 0$, $M(\vec{x})\rightarrow \left[ \bar{\rho} / \rho_R (\vec{x})\right]^p$ with $\bar{\rho}$ being the mean matter density of the Universe and $\rho_R(\vec{x})$ the density inside a sphere of radius $R$ around $\vec{x}$. If $p>0$ the mark gives more weight (and therefore more importance) to points that are in underdense regions, while if $p<0$ points in overdensities are weighted more. One can adjust these parameters to obtain different types of marks, that can weight in different ways the various components of the large-scale structure.

The marked power spectra are computed as follows. Firstly, the smoothed density field $\delta_R(\vec{x})$ is calculated on the vertex of a grid; the values of $\delta_R$ at the position of each matter particle are then computed via interpolation and a mark is assigned to each particle. Secondly, the marked power spectrum is computed as a power spectrum with each particle weighted by its mark.

We consider $5$ different values for each of the three mark parameters: $R=5,10,15,20,30\, h^{-1}$Mpc, $p=-1,0.5,1,2,3$ and $\delta_s = 0,0.25,0.5,0.75,1$, giving a total of $125$ different mark models. In total, millions of marked power spectra are available in the \textsc{Quijote} simulations.

\subsection{Correlation functions}

We compute correlation functions for 1) total matter field and 2) CDM+baryons field (only for simulations with massive neutrinos). The correlation functions are computed at redshifts 0, 0.5, 1, 2, and 3, in both real- and redshift-space. In the same way as for the power spectrum, redshift-space distortions are placed along one cartesian axis and three correlation functions are computed, one for each axis.

The procedure we use to compute the correlation functions is as follows. First, we assign particle masses to a regular grid with $N^3$ cells using the Cloud-in-Cell (CIC) mass assignment scheme and compute the density contrast: $\delta(\vec{x})=\rho(\vec{x})/\bar{\rho}-1$. We then Fourier transform the density contrast field to get $\delta(\vec{k})$. Next, we compute the modulus of each Fourier mode, $|\delta(\vec{k})|^2$, and Fourier transform back that field. Finally, we compute the correlation function by averaging modes that fall within a given radius interval. In redshift-space, the quadrupole and hexadecapole are computed in the same way as the monopole by weighing each mode by the contribution of the corresponding Bessel function.

By default we set $N$ to be equal to the cubic root of the number of particles in the simulation, but we also compute correlation functions in finer grids. In total, we provide over 1 million correlation functions.

\subsection{Bispectra}

We compute bispectra for the total matter field as well as for halo catalogs in both real- and redshift-space at redshifts 0, 0.5, 1, 2 and 3. We use a Fast Fourier Transform (FFT) based estimator similar to the estimators described
in~\cite{sefusatti2005}, \cite{scoccimarro2015},  and  \cite{sefusatti2016}.
We first interpolate matter particles/halos to a grid to compute the density contrast field, $\delta({\vec{x}})$,
using a fourth-order interpolation to get interlaced grids and then Fourier
transform the grid to get $\delta(\vec{k})$. We then measure the bispectrum monopole using
\begin{align}
\begin{split}
B_0(k_1,k_2, k_3) &= \frac{1}{V_B} \int\limits_{k_1}{\rm d}^3q_1 \int\limits_{k_2}{\rm d}^3q_2 \int\limits_{k_3}{\rm d}^3q_3~\delta_{\rm D}({\bm q_{123}})\\
&\qquad\times\delta({\bm q_1})~\delta({\bm q_2})~\delta({\bm q_3}) - B^{\rm SN}
\end{split}
\end{align}
where $\delta_{\rm D}$ is the Dirac delta function, $V_B$ is a normalization
factor proportional to the number of triplets in the triangle bin defined by
$k_1$, $k_2$, and $k_3$, and $B^{\rm SN}$ is the correction for Poisson shot
noise. To evaluate the integral, we take advantage of the plane-wave
representation of $\delta_{\rm D}$. For more details, we refer readers to \cite{Chang_19}\footnote{The code that we use to
evaluate $B_0$ is publicly available at \url{https://github.com/changhoonhahn/pySpectrum}}.
We use $\delta(\vec{x})$ grids with $N_{\rm grid} = 360$
and triangle configurations defined by  $k_1$, $k_2$, and $k_3$ bins of width
$\Delta k = 3 k_f = 0.01885~h{\rm Mpc}^{-1}$. For $k_{\rm max}=0.5~h{\rm Mpc}^{-1}$
there are 1898 triangle configurations. Redshift-space distortions are imposed along one Cartesian axis, same as the power spectrum, so we measure three bispectra, one for each axis.

In total, the \textsc{Quijote} simulations provide over 1 million bispectra.

\subsection{PDFs}
\label{subsec:PDF}

We estimate the probability density functions (PDF) of the matter, CDM+baryons and halo field in all the simulations at all redshifts. The PDFs are computed as follows. First, we deposit particle masses (or halo positions) to a regular grid with $N^3$ cells using the Cloud-in-Cell (CIC) mass assignment scheme. We then smooth that field with a Gaussian filter of radius, $R$. Finally, the PDF is calculated by computing the fraction of cells whose overdensity lie within a given interval. We compute the PDFs for many different values of $R$. By default we take $N$ to be the cubic root of the number of CDM particles in the simulation. In total, the \textsc{Quijote} simulations provide more than 1 million PDFs.

\section{Applications}
\label{sec:Applications}

The \textsc{Quijote} simulations have been designed to address two main goals: 1) to quantify the information content on cosmological observables, and 2) to provide enough statistics to train machine learning algorithms. In this section we describe a few examples of applications of the simulations.

\subsection{Information content from observables}
\label{subsec:Information_content}

\begin{figure*}
\begin{center}
\includegraphics[width=1.0\textwidth]{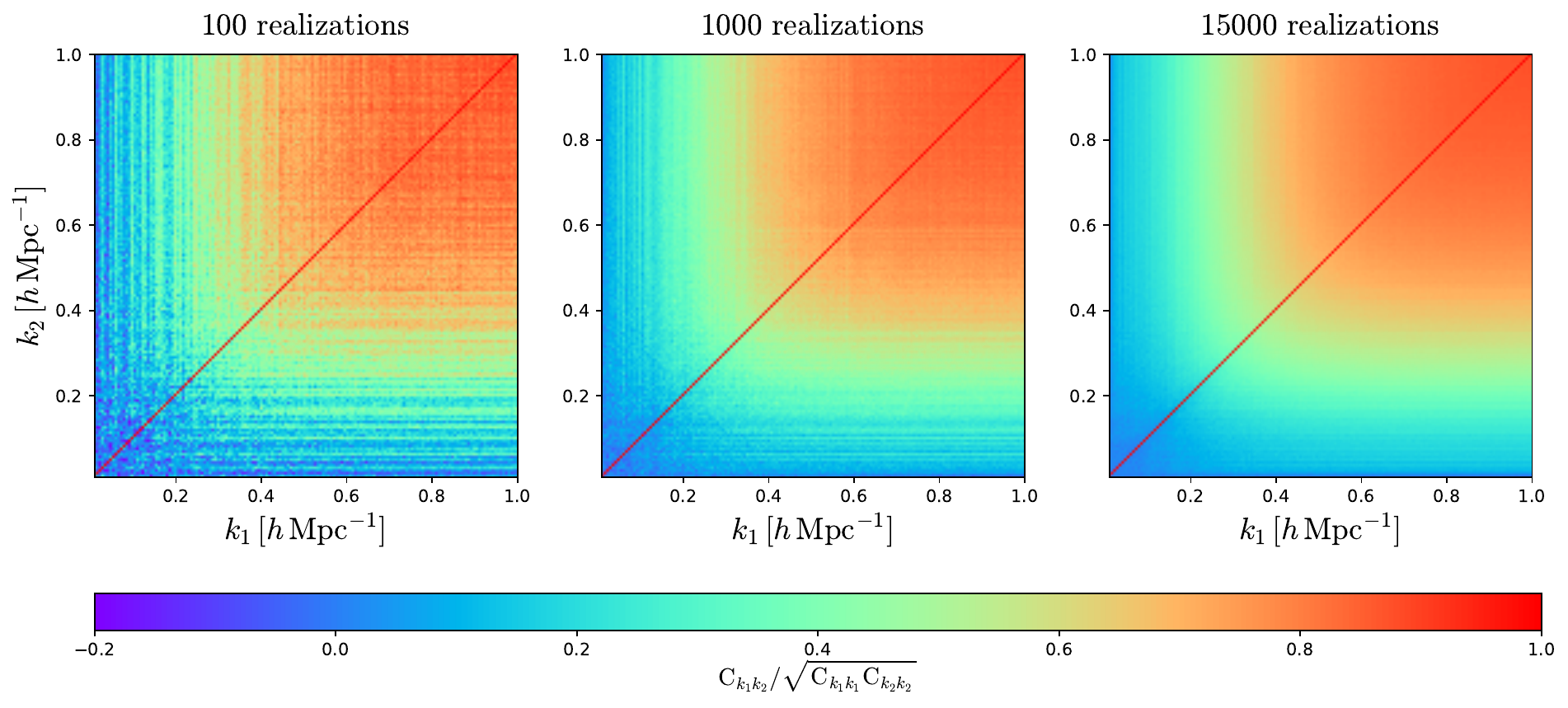}
\caption{Correlation matrix of the matter power spectrum at $z=0$ computed using 100 (left), 1000 (center), and 15000 (right) realizations. On large-scales, modes are decoupled, so correlations are small. On small scales, modes are tightly coupled, and the amplitude of the correlation is high. As expected, the noise in the covariance matrix shrinks with the number of realizations.}
\label{fig:Covariance}
\end{center}
\end{figure*}
\begin{figure*}
\begin{center}
\includegraphics[width=1.0\textwidth]{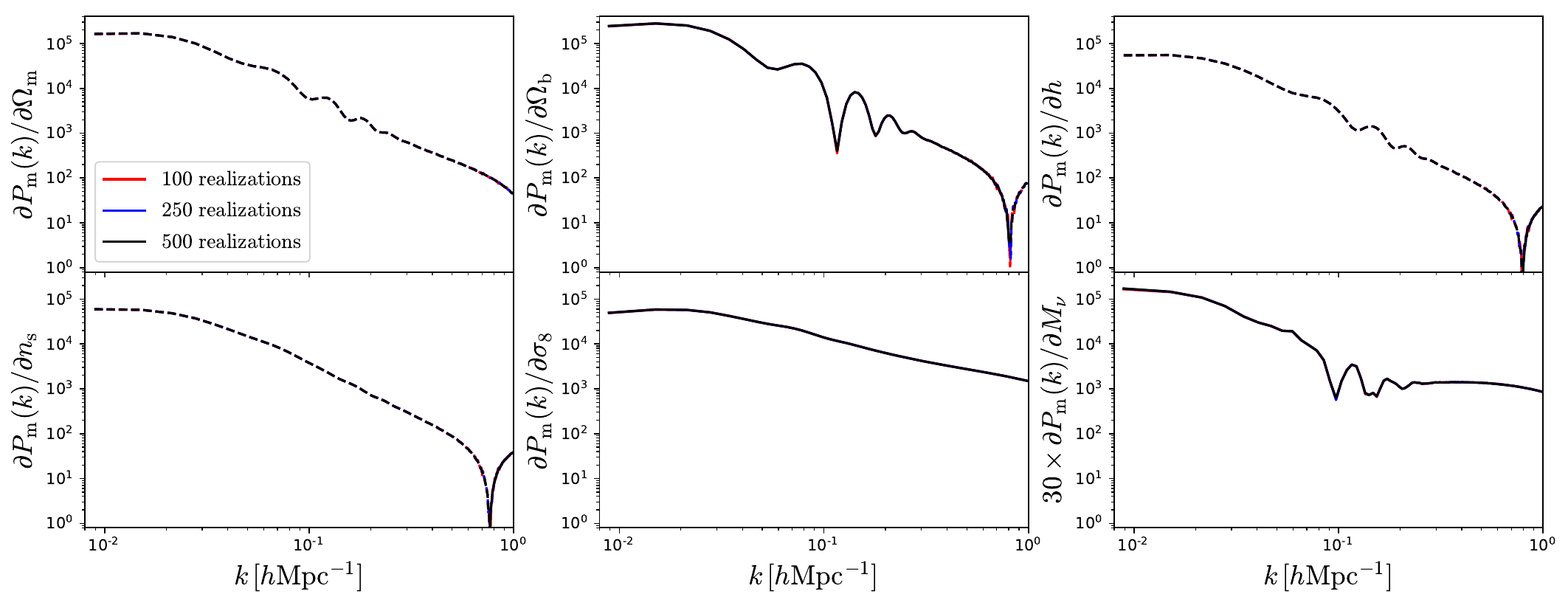}
\caption{Derivatives of the matter power spectrum in real-space with respect to $\Omega_{\rm m}$ (upper-left), $\Omega_{\rm b}$ (upper-middle), $h$ (upper-right), $n_s$ (bottom-left), $\sigma_8$ (bottom-middle), and $M_\nu$ (bottom-right) at $z=0$. Solid and dashed lines represent positive and negative values of the derivatives, respectively. We show the derivatives when computed using 100 (red), 250 (blue), and 500 (black) realizations. It can be seen how results are very converged against the number of realizations.}
\label{fig:derivatives}
\end{center}
\end{figure*}

As discussed in the introduction, it is currently unknown what statistic or statistics should be used to retrieve the maximum cosmological information from non-Gaussian density fields. One way to quantify the information content on a set of cosmological parameters, $\vec{\theta}$, given a statistics $\vec{S}$, is through the Fisher matrix formalism. The Fisher matrix is defined as
\be
F_{ij}=\sum_{\alpha,\beta}\frac{\partial S_\alpha}{\partial \theta_i}C^{-1}_{\alpha \beta}\frac{\partial S_\beta}{\partial \theta_j}~,
\label{eq:Fisher}
\ee
where $S_i$ is the element $i$ of the statistic $\vec{S}$ and $C$ is the covariance matrix
\be
C_{\alpha \beta} = \langle (S_\alpha-\bar{S}_\alpha)(S_\beta - \bar{S}_\beta) \rangle; \hspace{1cm} \bar{S}_i = \langle S_i \rangle~.
\ee
Notice that in Eq. \ref{eq:Fisher} we have set to 0 the term \citep[see e.g.][]{Tegmark_96}
\be
\frac{1}{2}{\rm Tr} \left[C^{-1}\frac{\partial C}{\partial \theta_\alpha} C^{-1} \frac{\partial C}{\partial \theta_\beta}\right]~.
\ee
This is because this term is expected to be small \citep{Kodwani_2019} but including it will also lead to an underestimation of the parameter errors \citep{Carron_2013, Justin_2018}.

The error on the parameter $\theta_i$, marginalized over the other parameters, is given by
\be
\delta\theta_i\geq \sqrt{\left(F^{-1}\right)_{ii}}~.
\ee
Thus, in order to quantify the constraints that a given statistic can place on the value of the cosmological parameters we only need two ingredients: 1) the covariance matrix of the statistic(s) and 2) the derivatives of the statistic(s) with respect to the cosmological observables. As discussed in detail in Sec. \ref{sec:Simulations}, the \textsc{Quijote} simulations have been designed to numerically evaluate those two pieces.

In this paper we consider one of the simplest applications of our simulations: the information content on the matter power spectrum. In Fig. \ref{fig:Covariance} we plot the correlation matrix of the matter power spectrum at $z=0$, defined as

\begin{equation}
\frac{C_{k_i k_j}}{\sqrt{C_{k_i k_i}C_{k_j k_j}}}
\end{equation}
when computed using 100 (left), 1000 (middle) and 15000 (right) realizations of the fiducial cosmology. As can be seen, results are noisy when computing the covariance with few realizations; this in turn, affects the results of the Fisher matrix analysis.

As it is well-known, on large-scales, the different Fourier modes are decoupled, and the covariance matrix is almost diagonal. On small-scales, modes with different wavenumbers are coupled, giving rise to non-diagonal elements, whose amplitude increases on smaller scales. Notice that previous works have investigated in detail the properties of the covariance matrix using a very large set of simulations \citep{Blot_2015, Blot_2016}.

In Fig. \ref{fig:derivatives} we show the second ingredient we need to evaluate the Fisher matrix: the partial derivatives of the matter power spectrum with respect to the cosmological parameters. In our case, we only consider $\Omega_{\rm m}$, $\Omega_{\rm b}$, $h$, $n_s$, $\sigma_8$, and $M_\nu$, and show results at $z=0$. In that Figure we show the derivatives when computed using different number of realizations. It can be seen how results are well converged, all the way to $k=1~h{\rm Mpc}^{-1}$. We can also see how the derivatives are different among parameters, pointing out that the matter power spectrum alone can provide information on each parameter separately.

\begin{figure*}
\begin{center}
\includegraphics[width=0.99\textwidth]{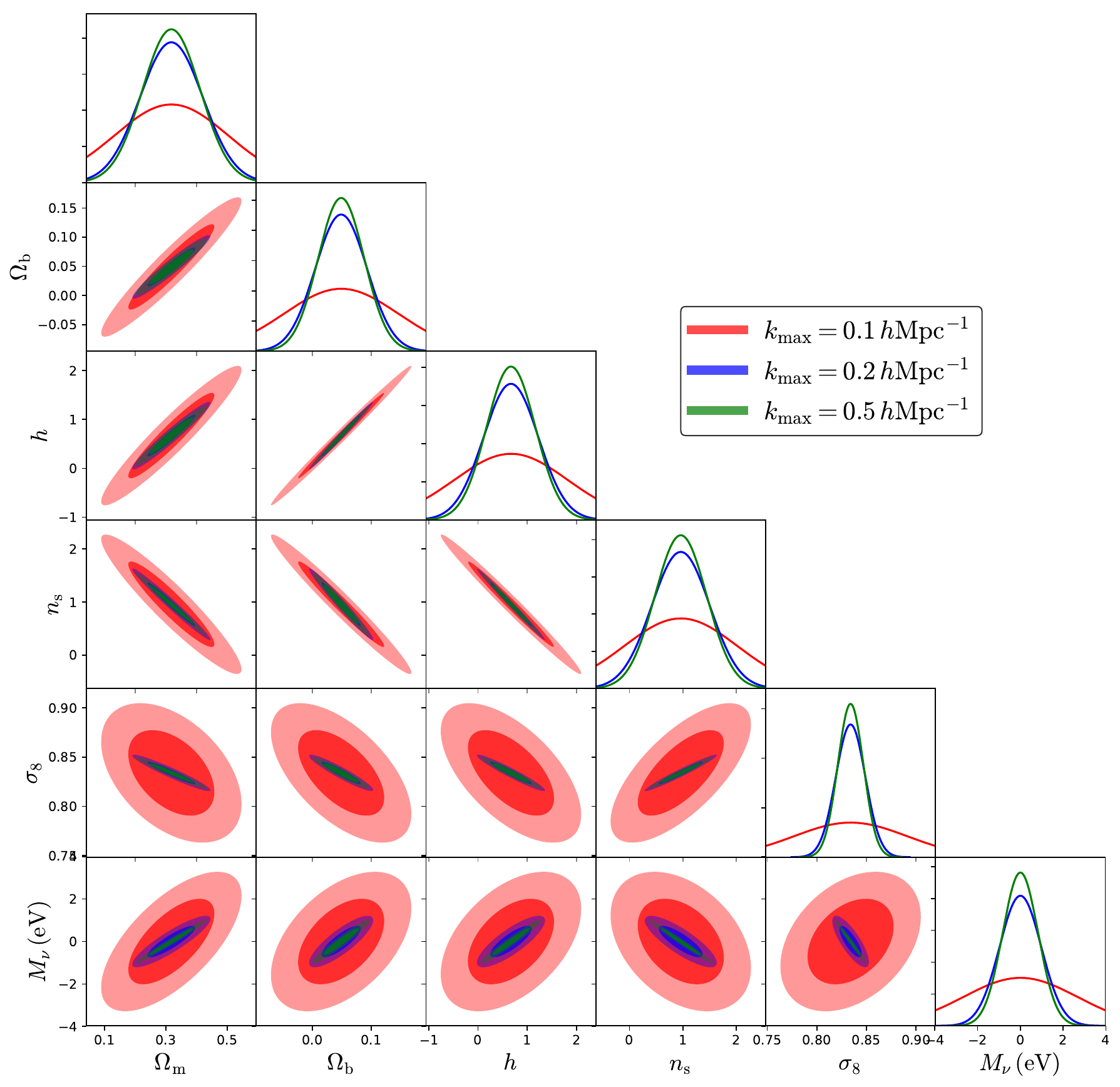}
\caption{Constraints on the value of the cosmological parameters from the matter power spectrum in real-space at $z=0$ for $k_{\rm max}=0.2$ (red), 0.5 (blue) and 1.0 (green) $h{\rm Mpc}^{-1}$. The small and big ellipses represent the $1\sigma$ and $2\sigma$ constraints, respectively. The panels with the solid lines represent the probability distribution function of each parameter. As we move to smaller scales, the constraints on the parameters improve. On the other hand, the fact that modes on small-scales are highly coupled, limit the amount of information that can be extracted from the matter power spectrum by going to smaller scales.}
\label{fig:Pk_info}
\end{center}
\end{figure*}

With the covariance matrix and the derivatives we can evaluate the Fisher matrix and determine the constraints on the cosmological parameters. We have verified that our results are converged, i.e. constraints do not change if the covariance and derivatives are evaluated with less realizations. We have also checked that our results are robust against different evaluations of the neutrino derivatives. We show the results on Fig. \ref{fig:Pk_info} when we consider the matter power spectrum down to $k_{\rm max}=0.1$ (red), 0.2 (blue) and 0.5 (green) $h{\rm Mpc}^{-1}$. As expected, the smaller the scales, the more cosmological information we can extract and the tighter the constraints on the parameters. However, the gain with scale does not scale proportional to $k_{\rm max}^3$, as naively expected just by counting number of modes. There are two main reasons for this behaviour: 1) the covariance becomes non-diagonal on small scales; modes become correlated and therefore the number of independent modes do not scale as $k_{\rm max}^3$, and 2) degeneracies among parameters limit the amount of information that can be extracted.

\begin{figure*}
\begin{center}
\includegraphics[width=0.99\textwidth]{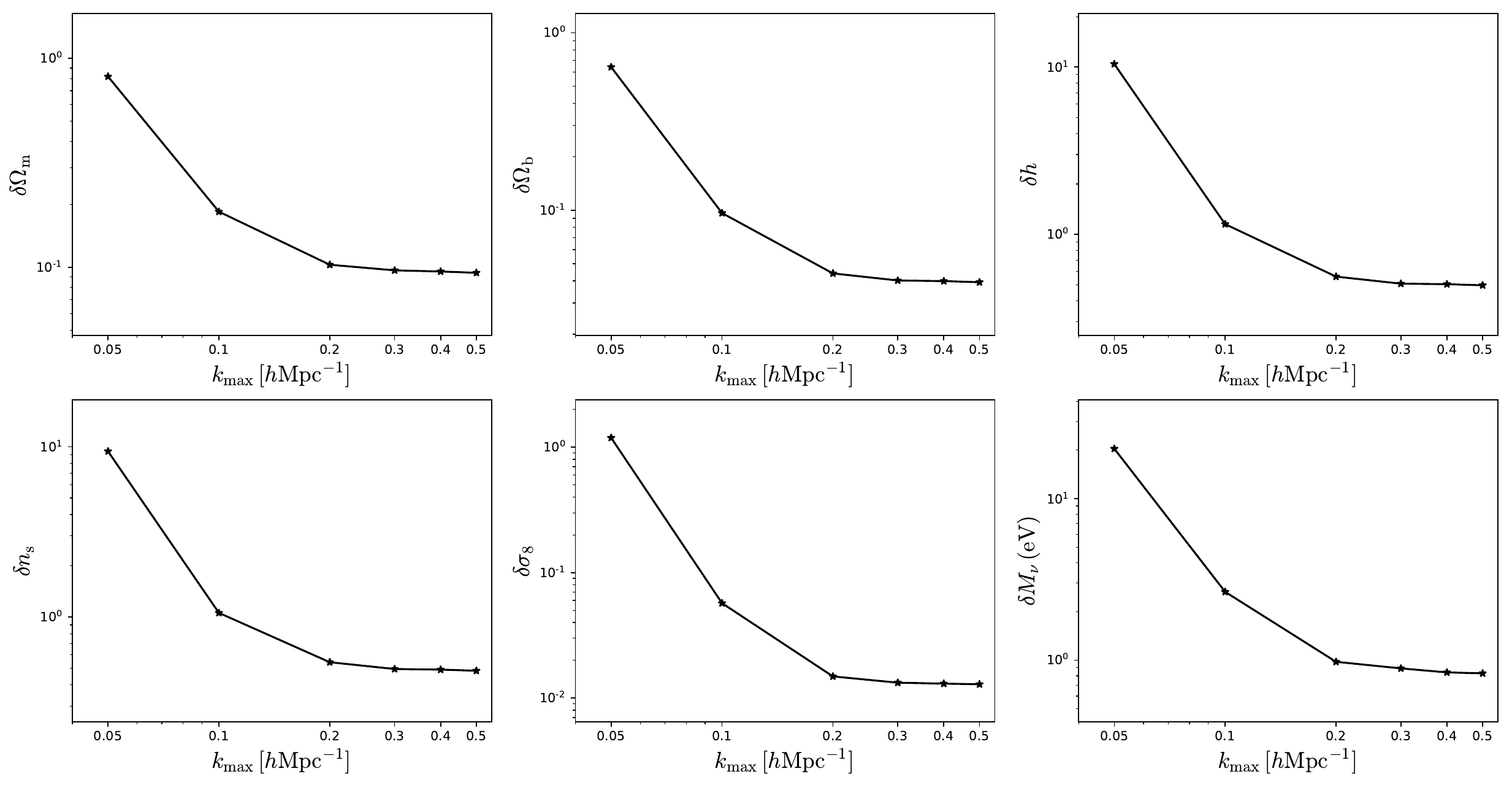}
\caption{Marginalized $1\sigma$ constraints on the value of the cosmological parameters from the matter power spectrum in real-space at $z=0$ as a function of $k_{\rm max}$. As we go to smaller scales, the information content on the different parameters tends to saturates. This effect is mainly driven by degeneracies among the parameters.}
\label{fig:Pk_info2}
\end{center}
\end{figure*}

In Fig. \ref{fig:Pk_info2} we show the marginalized $1\sigma$ constraints on the value of the cosmological parameters as a function of $k_{\rm max}$. As can be seen, the constraints on the parameters tend to saturate on small scales. We note that this result is mainly driven by degeneracies among parameters rather than the covariance becoming non-diagonal.

The cosmological information that was present on the matter power spectrum at high-redshift on small scales has now leaked into other statistics due to non-linear gravitational evolution. The \textsc{Quijote} simulations can be used to quantify it. The information content on the full halo bispectrum in redshift-space is estimated in \cite{Chang_19}. The constraints on the parameters by combining the power spectrum, halo mass function and void size function is presented in \cite{Paco_19}, while sensitivity of the cosmological parameters to the marked power spectrum is shown in \cite{Massara_19}. \cite{Cora_19} quantifies the information content on the PDF of the 3D matter field.

\subsection{Information content from neural nets}
\label{subsec:IMNN}

A  way of searching for new statistics is using information maximising neural networks (IMNN)~\citep{Charnock_2018}. The IMNN is designed to automatically find informative, non-linear summaries of the data. The method uses neural networks to transform non-Gaussian data in to the set of optimally compressed, Gaussianly-distributed summaries via maximisation of the Fisher information. These summaries can then be used in a likelihood-free inference setting or even directly as pseudo-maximum likelihood estimators of the parameters. By building neural networks using physically motivated principles, not only will we obtain informative summaries of the data, but we will be able to attribute these summaries to real space effects, hence learning even more about the connection between data and the underlying cosmological model. As an input, the IMNN requires simulated data to compute the covariance of the summaries and the derivative of the summaries with respect to model parameters. The design of the \textsc{Quijote} simulations enables this novel approach to identify and quantify information content from new observables.

\subsection{Likelihood-free inference}
\label{subsec:ML}

Besides quantifying the information content on cosmological observables, the \textsc{Quijote} simulations have been designed to provide enough data to train machine learning algorithms. In this subsection we present a very simple application using a well-known machine learning algorithm: the random forest.

We use the 2000 standard simulations of the LH latin-hypercube run at fiducial resolution. For each simulation, we compute the 1-dimension PDF when the density field is smoothed on a scale of $5~h^{-1}{\rm Mpc}$ using a top-hat filter (see subsection \ref{subsec:PDF} for further details) at $z=0$. For each simulation we have thus an input, the value of the PDF on a set of overdensity bins, and a label, the value of the 5 cosmological parameters that we vary on those simulations: $\Omega_{\rm m}$, $\Omega_{\rm b}$, $h$, $n_s$, and $\sigma_8$. Our purpose, is to find the function that map these two vectors, i.e.
\begin{equation}
\vec{\theta} = f(\vec{{\rm PDF}}(1+\delta)).
\end{equation}

The standard way to find the function $f$ is to develop a theoretical model that outputs the PDF for a given value of the cosmological parameters \citep{Cora1, Cora2, Cora3, Gruen_18}. A different approach is to identify features in the data that can be used as a link to the value of the labels. In our case, we search features on the input data using a simple random forest regressor.

We split our data into two sets: 1) a training set with the results of 1600 simulations and 2) a test set with the remaining 400 simulations. We train the random forest algorithm using the input and output of the training set. We then use the trained random forest, to predict the value of the cosmological parameters from the PDF of the simulations of the test set. We emphasize that the random forest has never seen the data from the test set, and therefore, the output from the test set is a true prediction.

\begin{figure*}
\begin{center}
\includegraphics[width=0.99\textwidth]{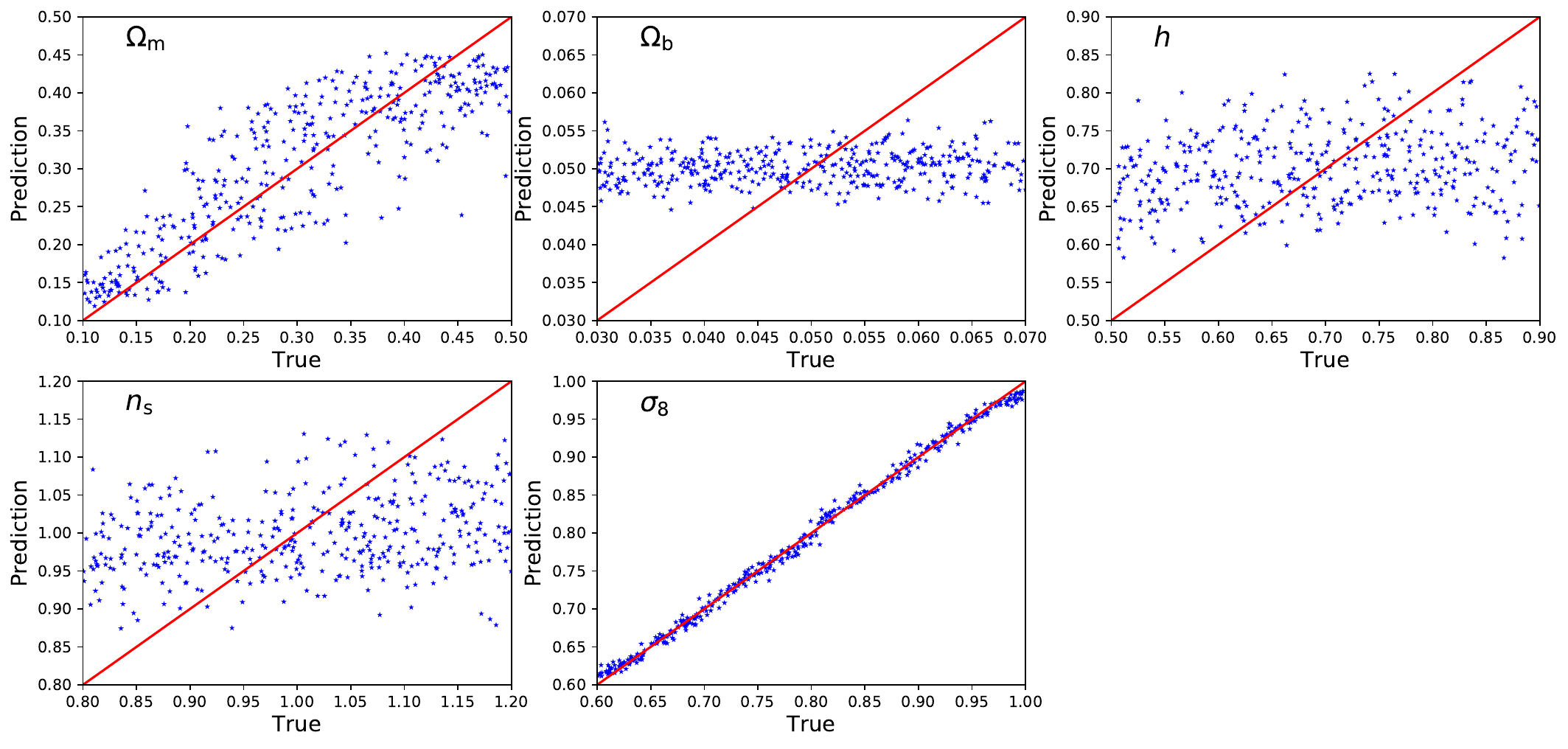}
\caption{For each of the 2000 standard simulations at fiducial-resolution in the LH hypercube we have measured the 1-point PDF of the matter density field smoothed on a scale of $5~h^{-1}{\rm Mpc}$. We have then split the data into two different sets: 1) traininig set (1600 simulations) and 2) test set (400 simulations). We have trained a random forest algorithm to find the mapping between the measured values of the 1-pt PDF and the value of the cosmological parameters using the training set. Once trained, we have used the test set (that the algorithm has never seen) to see how well we can predict the cosmological parameters from unlabeled PDF measurements. Each panel shows the predicted value versus the true one for $\Omega_{\rm m}$ (top-left), $\Omega_{\rm b}$ (top-middle), $h$ (top-right), $n_s$ (bottom-left), and $\sigma_8$ (bottom-middle). We find that the random forest can only predict the value of $\sigma_8$ and $\Omega_{\rm m}$ from the PDF. We emphasize that no theory model/template has been used to relate the PDF with the value of the parameters.}
\label{fig:random_forest}
\end{center}
\end{figure*}

We show the results on this exercise in Fig. \ref{fig:random_forest}. In each panel, the y-axis represents the prediction of the random forest, while the x-axis is the true value. As can be seen, the random forest learns how to accurately predict the value of $\sigma_8$ from the fully non-linear PDF without need of developing a theory model. Another parameter that the random forest is able to predict is $\Omega_{\rm m}$, although less accurately. $\Omega_{\rm b}$, $h$ and $n_s$ are however unconstrained by the random forest; failing to capture the parameter dependence, the random forest regressor minimizes the training loss by outputting values close to the mean of the training set. Notice that it is physically expected that for a volume of $1~(h^{-1}{\rm Gpc})^3$, and only using the 1D PDF at a single smoothing scale, the constraints on those parameters will not be very tight.

It is however possible to improve these results by identifying features in the 3-dimensional density field, instead of on summary statistics. For instance, \cite{Ana_19b} uses convolutional neural nets to identify features that allow constraining the value of the cosmological parameter directly from the 3D density field of the \textsc{Quijote} simulations.

\subsection{New non-Gaussian statistics}
\label{subsec:WST&PH}

In previous years, it has been shown that particular low-variance representations inspired from deep neural networks can efficiently characterize non-Gaussian fields. Based on the multi-scale decomposition achieved by the wavelet transform, these representations are built from successive applications of the so-called scattering operator on the field under study (convolution by a wavelet followed by a modulus operator,~\cite{mallat2012group}), and/or from phase harmonics of its wavelet coefficients (multiplication of their phase by an integer,~\cite{mallat2018phase}). They can then be analyzed directly as well as from their covariance matrix, and have obtained state-of-the-art classification results when applied to handwritten and texture discrimination~\citep{bruna2013invariant,sifre2013rotation}.

\begin{figure*}
\begin{center}
\includegraphics[width=0.99\textwidth]{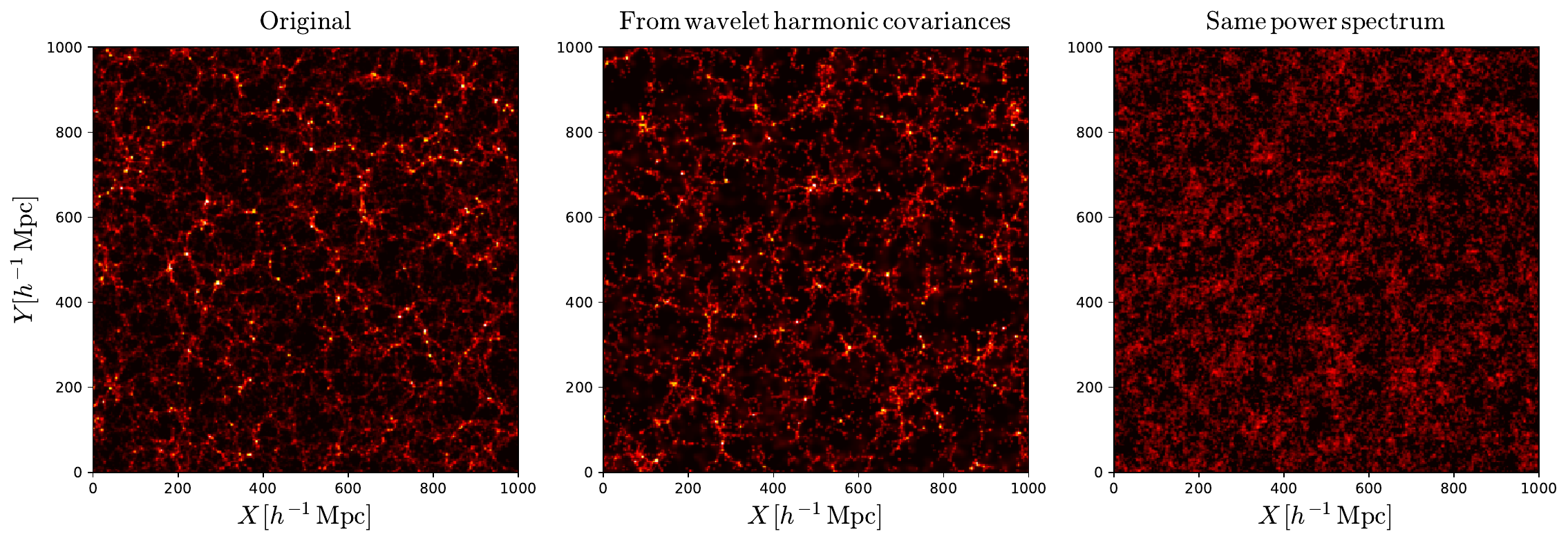}
\caption{Left: Projected density field of a $1000\times 1000 \times 60~(h^{-1}{\rm Mpc})^3$ region at $z=0$ for a fiducial cosmology. Center: Simulated field with the same covariance coefficients of wavelet harmonics than the original one (around one thousand coefficients). Right: Gaussian field of same power spectrum than the original one. All these images have a resolution of $256\times26$ pixels. One sees that in contrast to the power spectrum, the phase harmonic coefficients are efficient to extract statistical features from an image.}
\label{fig:synthesis}
\end{center}
\end{figure*}

The use of tailored representations to comprehensively characterize non-Gaussian fields has several advantages with respect to what can be achieved with deep neural networks. Indeed, as the structure of these representations are given and do not necessitate any training stage, they open a path to the interpretability of the results obtained~\citep{allys2019rwst}. For the same reasons, these statistical descriptions can be used even when no large amount of data is available, since they do not need any training to be constructed. This is illustrated by the ability to synthesize very good looking synthetic fields from only one given sample, see below.

An important application of these non-Gaussian representations is to model the statistics of cosmological observations, e.g. of a projected density field. This unsupervised learning problem amounts to estimate the probability distribution of such observations, which are stationary, given one or more sample. One can then generate new maps by sampling this distribution. Following standard statistical physics approaches, the probability distribution of Quijote simulations are modeled as a maximum entropy distribution conditioned by moments~\cite{bruna2018multiscale}. The main difficulty is to define appropriate moments which are sufficient to capture the statistics of the field. The right image in Fig.~\ref{fig:synthesis} was sampled from a Gaussian process, which is a maximum entropy process conditioned by second order moments. It thus has the same power spectrum as the original. The middle image in Fig.~\ref{fig:synthesis} was sampled from a nearly maximum entropy distribution conditioned by wavelet harmonic covariance coefficients~\citep{mallat2018phase}. One can observe from this figure that the image obtained from wavelet harmonic covariances captures better the statistics of the original, including the geometry of high amplitude outliers and filaments, although it uses fewer moments than the Gaussian model. Indeed, wavelet harmonic moments also depend upon the correlation of phases across scales, which are responsible for the creation of these outliers, whereas Gaussian fields have independent random phases.

A second application of these new statistical descriptors is to infer relevant physical parameters from cosmological observations, which is the goal of ongoing works. Such results would then be compared as a benchmark to the results obtained using standard statistics as e.g. the power spectrum or the bispectrum. The \textsc{Quijote} simulations being designed to quantify the information content on cosmological observables, they form an ideal set of data for this purpose.

\subsection{Forward modeling \& simulation emulators}
\label{subsec:forward_model}

The relatively high-resolution\footnote{This statement applies in comparison with the simulations used in \cite{He_19}.} and large parameter space of the \textsc{Quijote}
simulation suites enable us to build more accurate machine learning models of
the structure formation.
\cite{He_19} showed that the highly nonlinear structure
formation process, simulated with particle-mesh (PM) gravity
solver~\citep{FastPM} with fixed cosmological parameters, can be emulated with
convolutional neural networks (CNNs).
The CNN model is trained to predict the simulation outputs given their initial
conditions (linearly extrapolated to the target redshift).
Its accuracy is comparable to that of the training simulations and much more
than that of 2LPT commonly used to generate galaxy mocks
\citep[e.g.][]{PTHALOS}, while at a much lower computation cost.
The gain in both accuracy and efficiency proves machine learning a promising
forward model of the Universe.

\begin{figure}
\begin{center}
\includegraphics[width=0.49\textwidth]{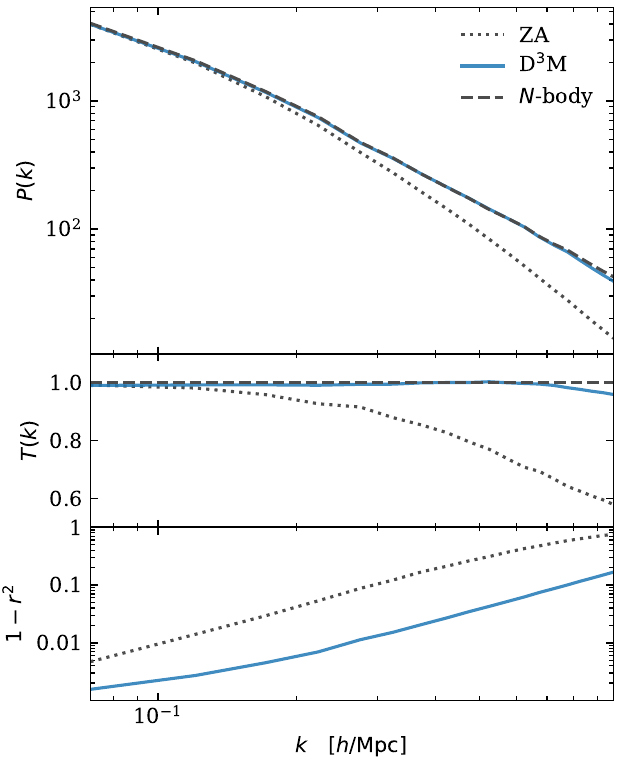}
\caption{The top panel shows the power spectra $P(k)$ predicted by Zeldovich
approximation (grey dotted), Quijote simulation (grey dashed), and CNN
model dubbed D$^3$M (blue solid).
The middle panel shows the transfer functions $T$ – the square root of the
ratio of each predicted power spectrum to that of the Quijote simulation (as
the ground truth).
In the bottom row we show the fraction of variance that cannot be explained by
each model, by the quantity $1 - r^2$ where $r$ is the correlation coefficient
between the predicted fields and the true fields.
$T$ and $r$ captures the quality of the model predictions.
As $T$ and $r$ approach one, the model prediction asymptotes to the ground
truth \citep{He_19}.
On both benchmarks the D$^3$M predictions are nearly perfect from linear to
nonlinear scales.}
\label{fig:ZA2Nbody}
\end{center}
\end{figure}

The \textsc{Quijote} suite are full $N$-body simulations that resolve gravity
to smaller scales than a PM solver.
This enables training more accurate CNN models deeper into the nonlinear
regime. Fig.~\ref{fig:ZA2Nbody} presents such an example that the machine learning
model from \cite{He_19} can make accurate predictions once trained with the
Quijote data.

Furthermore, with the set of latin-hypercube simulations (see Sec.~\ref{subsec:LH}), we are able to train CNN models that depend on chosen parameters in addition to the initial
conditions. This allows us to build an emulator at the field level.
Most of the existing emulators \citep[e.g.][]{Coyote,EuclidEmu,Aemulus2,Aemulus3,Aemulus4,Nishimichi_2018,Wibking:2017slg} are aimed mainly at predicting the
ensemble averaged 2-point statistics and halo abundance.
A CNN model conditional on cosmological parameters will open up the
opportunity to fully exploit the information encoded in the higher-order
statistics of the field.

\subsection{Super-resolution simulations}
\label{subsec:superresolution}

\begin{figure}
\begin{center}
\includegraphics[width=\hsize]{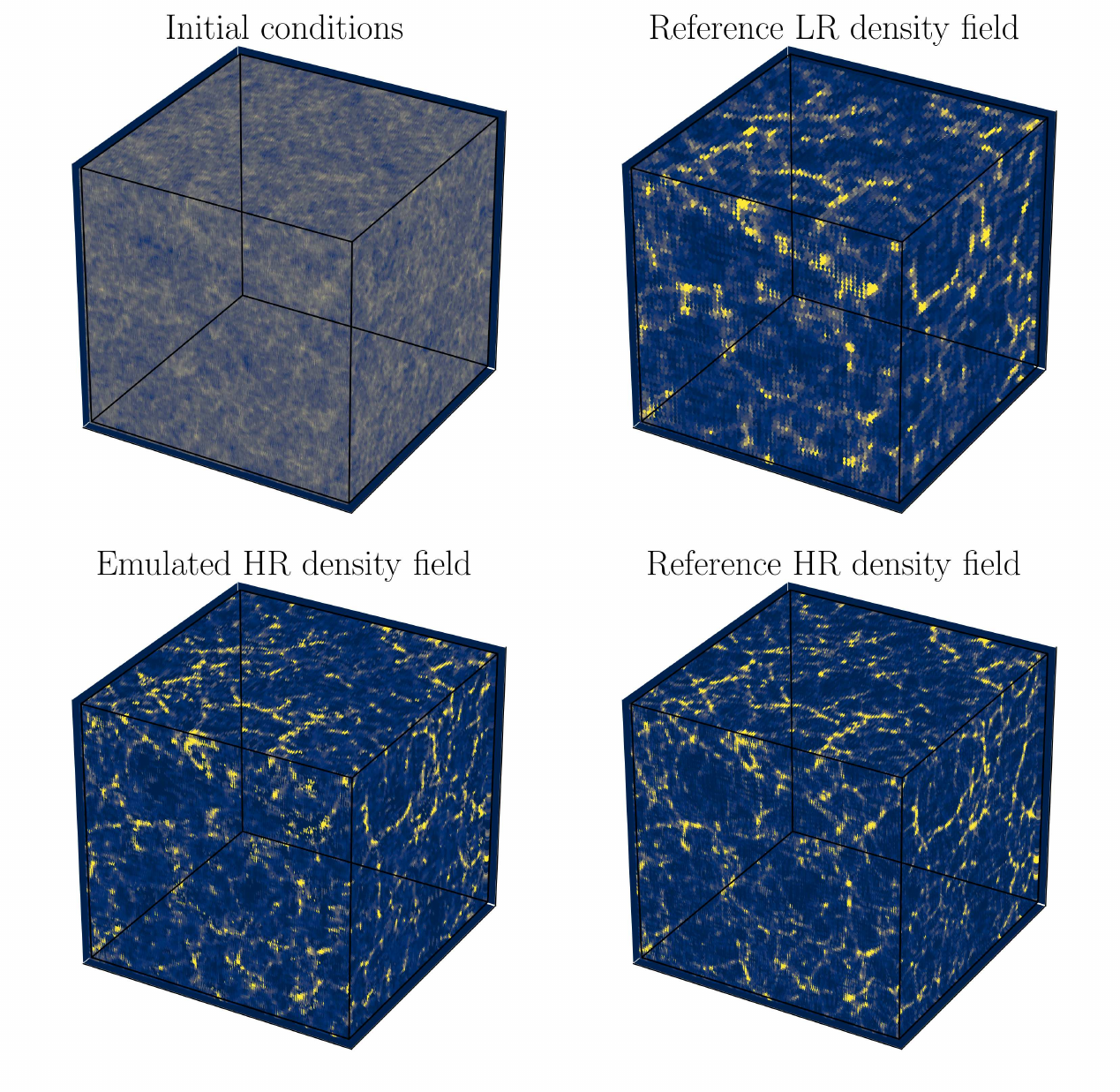}
\caption{Example of how to increase the resolution of a simulation using deep learning via the super-resolution emulator \citep{Ramanah_2020}. We combine the high-resolution initial conditions (top-left) with the $z=0$ low-resolution snapshot (top-right) to emulate a high-resolution snapshot at $z=0$ (bottom-left). The reference high-resolution simulation at $z=0$ is shown in the bottom-right panel for comparison.}
\label{fig:superresolution}
\end{center}
\end{figure}

Using the large quantity of high quality data available using the \textsc{Quijote} simulations, we are able to find methods with which we can accurately paint high-resolution features from computationally cheaper low-resolution simulations \citep{Ramanah_2020}. This super-resolution emulator relies on using physically motivated networks~\citep{Ramanah_2019} to perform a mapping of the distribution of the low-resolution cosmological distribution to the space of the high-resolution small-scale structure. Since the information content of the high-resolution simulations is far greater than in the low-resolution simulations, we can use the information contained in the high-resolution initial conditions as a well constructed prior from which to draw the data to in-paint the small-scale structure with statistical properties that mimic those of the high-resolution training data. In Fig. \ref{fig:superresolution}, we show an example of the output of our super-resolution emulator and its comparison with the reference high-resolution simulation.

By using this approach, not only do we obtain high-resolution simulations at a low cost, we also are able to inspect the physical network to learn about how the large-scale modes affect the small scale structure in real-space.

\subsection{Mapping between simulations}
\label{subsec:mapping_sims}

It is possible to use machine learning algorithms to find the mapping between the positions of particles in simulations with different cosmologies. In this way, from one simulation with a given cosmology it is possible to get new simulations with different cosmologies. This can be very useful in order to densely sample the parameter space or to compute covariance matrices in different regions of the parameter space.

\begin{figure}
\begin{center}
\includegraphics[width=0.45\textwidth]{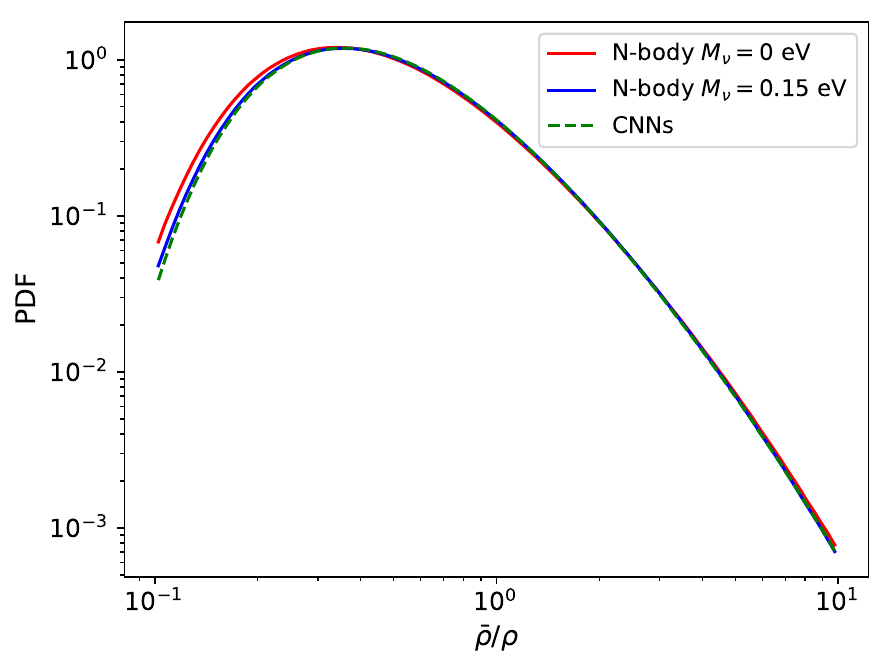}
\caption{The red and blue lines show the probability distribution function (PDF) of the CDM+baryons field for a cosmology with massless and massive neutrinos, respectively. We train neural networks to find the mapping between the massless and massive neutrino cosmologies. The dashed green line displays the PDF of the generated CDM+baryon field from the massless neutrino density field, showing a very good agreement with the expected blue line.}
\label{fig:map_sims}
\end{center}
\end{figure}

\cite{Giusarma_19} use deep convolutional neural networks to establish the link between the displacement field
\begin{equation}
\vec{d}_k = \vec{x}_{f,k} - \vec{x}_{i,k}
\end{equation}
where $\vec{x}_{f,k}$ and $\vec{x}_{i,k}$ are the final and initial position of particle $k$, in simulations with massless neutrinos and simulations with massive neutrinos \citep[see][for other methods to carry out this task]{Zennaro_19}. In Fig. \ref{fig:map_sims} we show an example of the results for a simple summary statistics: the 1D PDF.

\subsection{Statistical properties of paired fixed simulations}
\label{subsec:paired_fixed}

The large number of paired fixed simulations available in the \textsc{Quijote} simulations allow to investigate in detail their statistical properties. These simulations can save a lot of computational resources since they have been shown to largely reduce the amplitude of cosmic variance on certain statistics. Thus, they can be used to build emulators, evaluate likelihoods...etc.

\begin{figure}
\begin{center}
\includegraphics[width=0.45\textwidth]{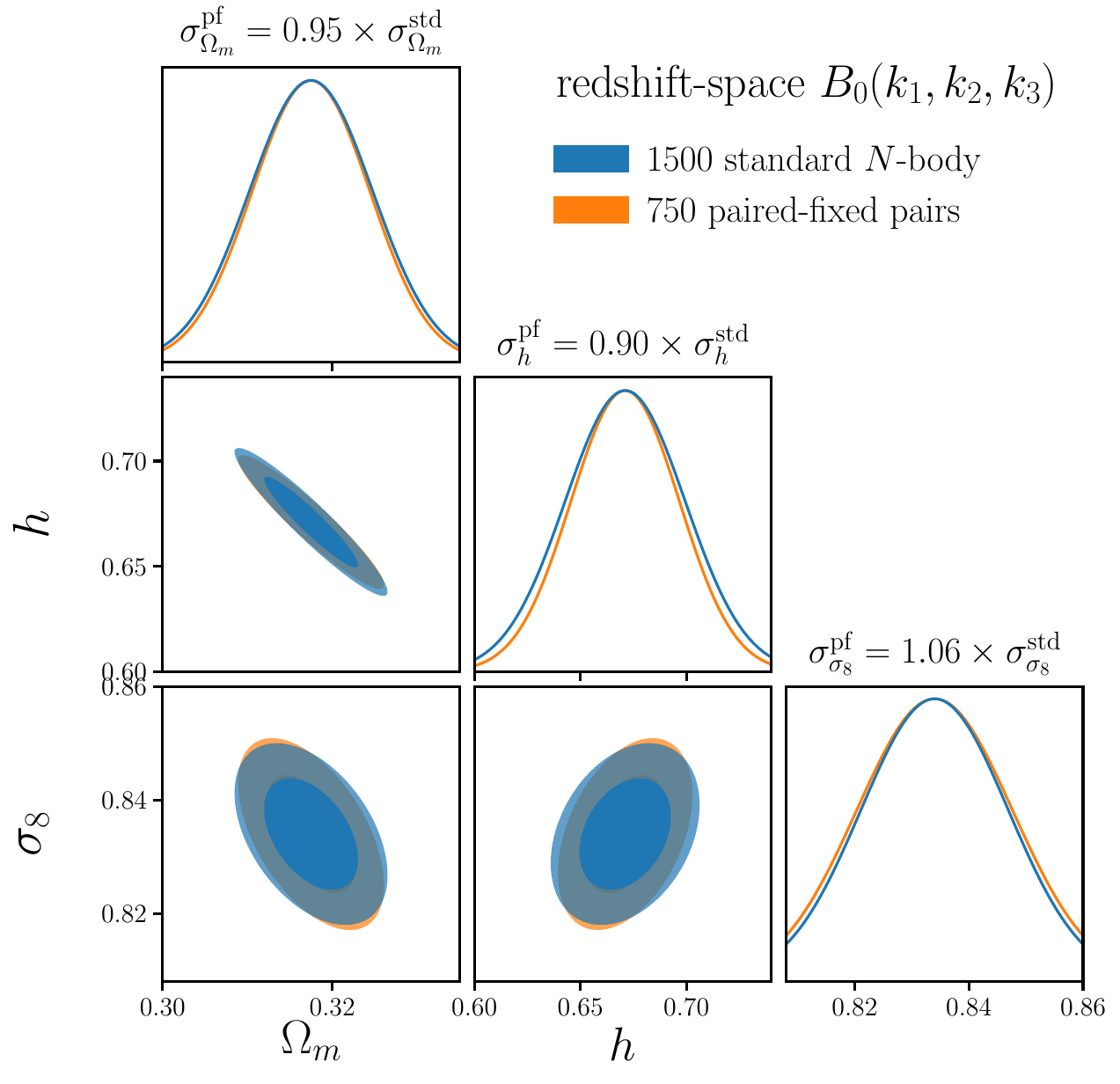}
\caption{We use the Fisher matrix formalism to quantify how accurately the
redshift-space halo bispectrum (down to $k_{\rm max}=0.5~h{\rm Mpc}^{-1}$)
can constrain $\Omega_{\rm m}$, $\sigma_8$ and $h$. In blue and orange we
show the results when the partial derivatives are computed using standard
and paired fixed simulations. We find that results are consistent and,
therefore, paired fixed simulations do not introduce a significant bias
for the halo bispectrum.}
\label{fig:Fisher_paired_fixed}
\end{center}
\end{figure}

\cite{Chang_19b} studies the impact of paired fixed simulations on the halo
bispectrum and performs a Fisher matrix analysis using both standard and
paired fixed simulations to evaluate the derivatives. They quantify how the
constraints on the cosmological parameters are affected by using standard
versus paired fixed simulations to evaluate the numerical derivatives. We
show some results for a subset of the parameters in
Fig.~\ref{fig:Fisher_paired_fixed}.

\section{Resolution tests}
\label{sec:resolution}

In this section we present some tests performed on the \textsc{Quijote} simulations to quantify the convergence of the simulations on several properties.

\subsection{Zel'dovich versus 2LPT}
\label{subsec:ZA_vs_2LPT}
The reason why we use the Zel'dovich approximation to generate initial conditions for cosmologies with massive neutrinos, and not 2LPT, is because, to our knowledge, it is unknown how to estimate the second-order scale-dependent growth factor and growth rate needed to use 2LPT in massive neutrino models.

\begin{figure*}
\begin{center}
\includegraphics[width=0.49\textwidth]{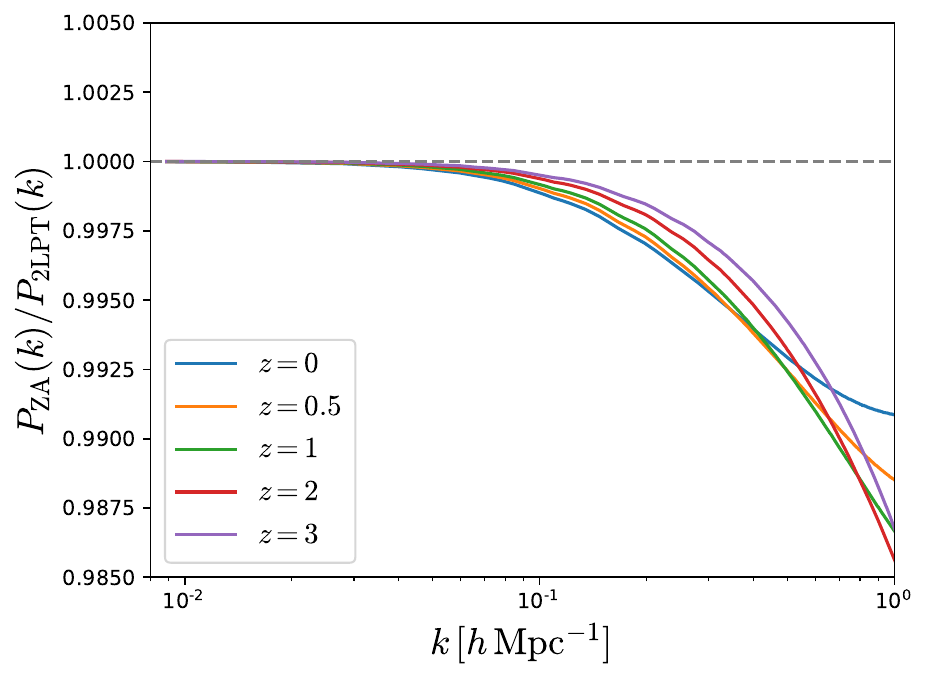}
\includegraphics[width=0.49\textwidth]{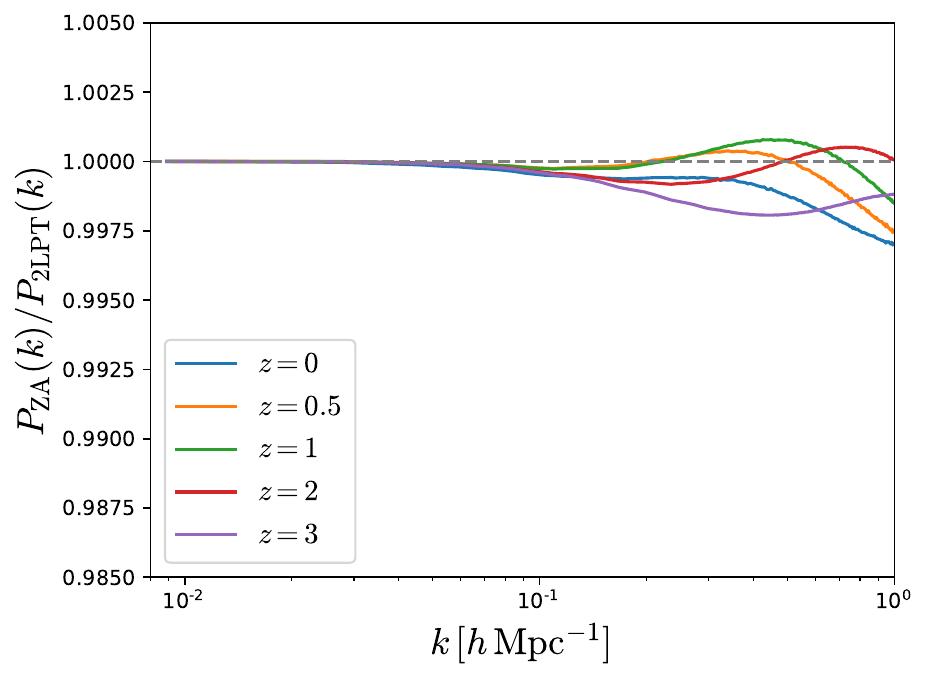}\\
\includegraphics[width=0.49\textwidth]{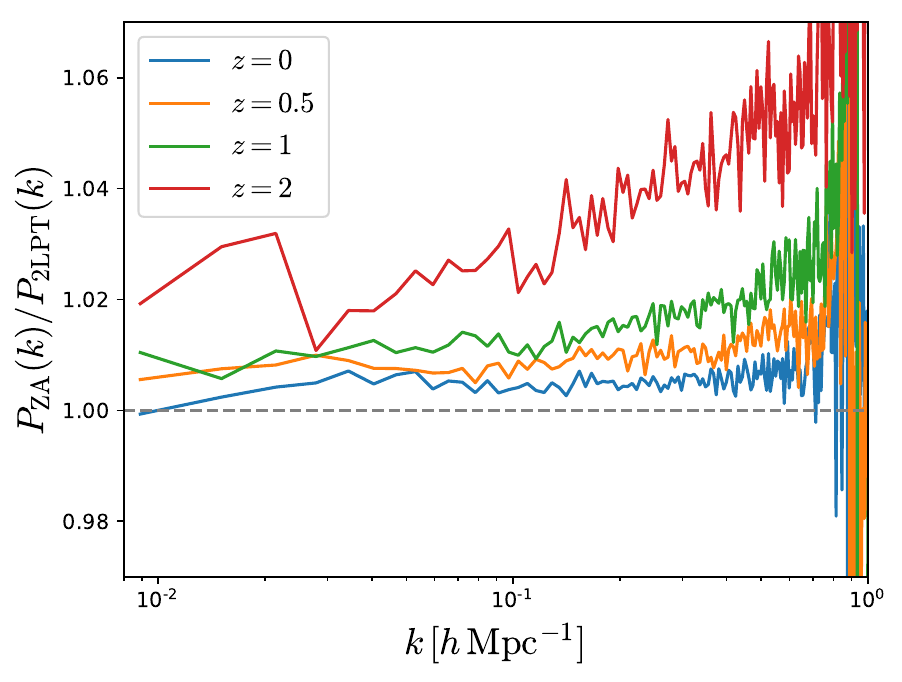}
\includegraphics[width=0.49\textwidth]{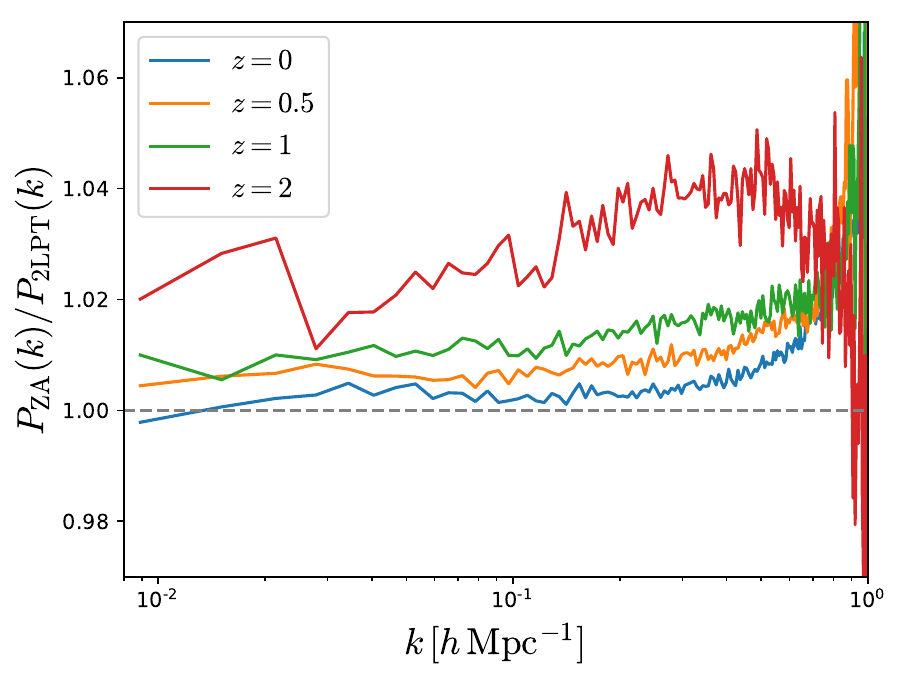}\\
\caption{Effect on the matter power spectrum in real- (top-left) and redshift-space (top-right) of generating the ICs using the Zel'dovich approximation versus 2LPT. The bottom panels show the same for the power spectrum of halos with masses above $3.2\times10^{13}~h^{-1}M_\odot$ in real- (bottom-left) and redshift-space (bottom-right). The plots show the ratio between the two power spectra as a function of wavenumber for different redshifts. For matter in real-space, the effect is below $1.5\%$, while in redshift-space the effect is below $0.5\%$ on all scales. For halos at low-redshift, the effect is $\lesssim1\%$. Near $k=1~h{\rm Mpc}^{-1}$, the halo power spectrum becomes negative (after subtracting shot-noise), and is severely affected by numerical noise.
Since the ICs of the massive neutrino simulations have been generated using the ZA, in comparison with 2LPT for the other models, it is important to keep in mind this effect when computing numerical derivatives.}
\label{fig:ZA_vs_2LPT}
\end{center}
\end{figure*}

Generating the initial conditions via Zel'dovich, instead of 2LPT, can induce small changes in the dynamics of the simulation particles, that can lead to small statistical differences \citep{Crocce_2006}. In order to quantify this effect, we have computed the matter and the halo power spectra (for halos with masses above $3.2\times10^{13}~h^{-1}M_\odot$) in 200 simulations of the fiducial cosmology: 100 simulations with Zeldovich ICs and 100 simulations with 2LPT ICs. The random seed are matched among the two sets. We show the results in Fig. \ref{fig:ZA_vs_2LPT}.

In the top panels we show the results for the matter power spectrum in real-space (top-left) and redshift-space (top-right). We find that differences in real-space are below $1.5\%$ at all the redshifts, while in redshift-space the effects are much smaller; below $0.25\%$. In the bottom panels we show the results for the power spectrum of halos in real-space (bottom-left) and redshift-space (bottom-right). We have corrected for Poissonian shot-noise by subtracting $1/\bar{n}$ to the measurements, where $\bar{n}$ is the number density of halos. Results at $z=3$ are very noisy, due to the very low number density of halos, thus, for clarity we do not show them. We find that differences in real- and redshift-space at low redshift are below $\simeq1\%$. The higher the redshift the larger the differences. The large variations we observe around $k=1~h{\rm Mpc}^{-1}$ are due to the halos power spectrum becoming very small, and therefore highly affected by numerical noise. Notice that at low-redshift, most of the differences we observe between the halos power spectra have a very mild scale-dependence. Thus, marginalizing over an overall amplitude can get rid of most of this effect.

We also carry out the above analysis for the bispectrum of halos in real- and redshift-space at $z=0$ down to $k_{\rm max}=0.5~h{\rm Mpc}^{-1}$. We find that differences in redshift-space can be around $10\%$, and slightly larger in real-space.

When making Fisher forecasts analysis, it is important to keep this effect in mind, as the additional scale-dependent present in the models with massive neutrinos may slightly affect the results. For this reason, when computing derivatives with respect to neutrino masses, we recommend using the simulations with Zel'dovich initial conditions from the fiducial model, instead of the 2LPT ones.

\begin{figure*}
\begin{center}
\includegraphics[width=1.0\textwidth]{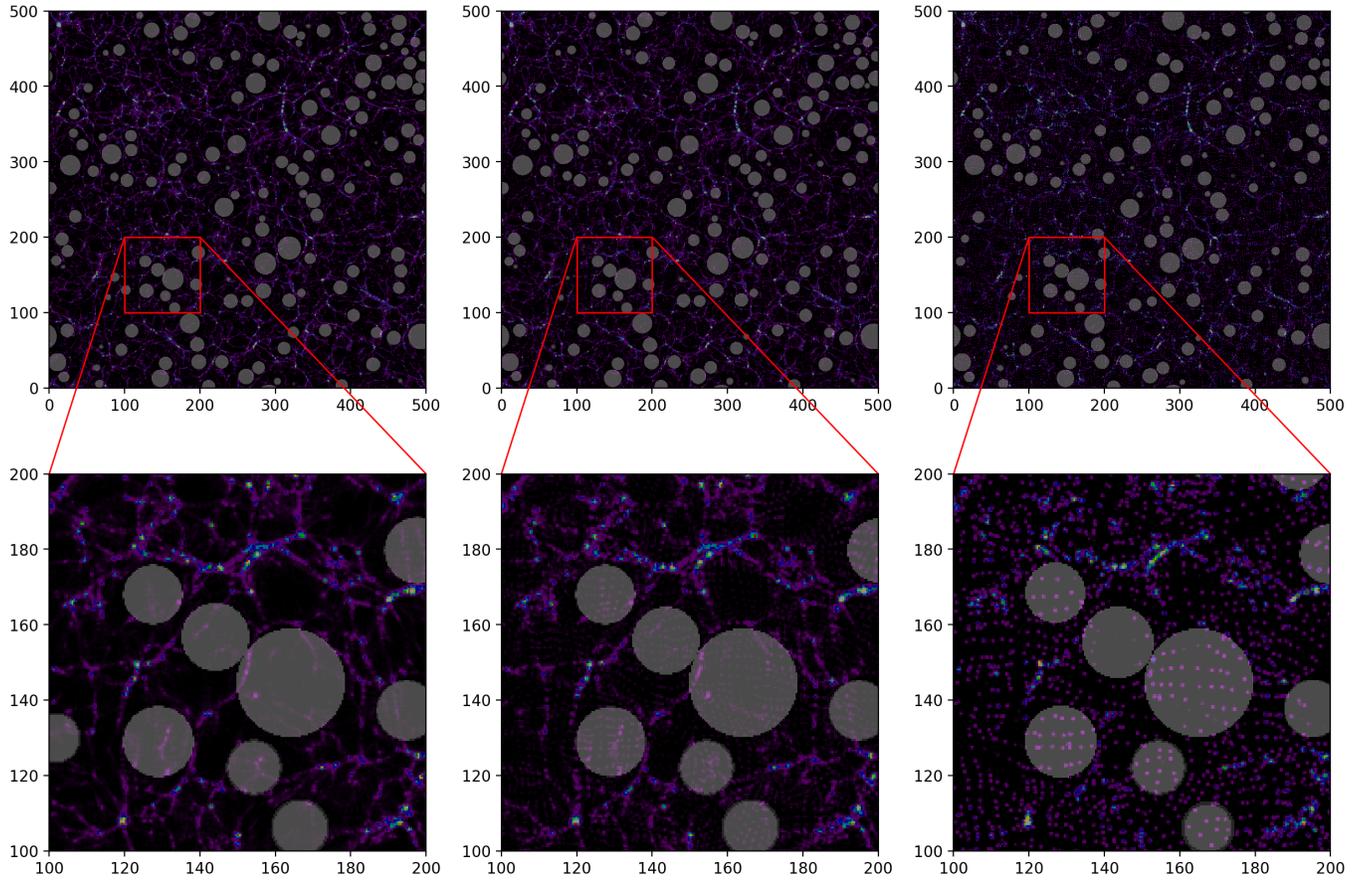}
\caption{We have identified voids (white spheres) in three simulations with the same random seed but different mass and spatial resolutions at $z=0$. As can be seen, our void finder is relatively robust against these changes, at least for the largest voids.}
\label{fig:void_finder}
\end{center}
\end{figure*}

\subsection{Clustering}

One important aspect to consider when analyzing numerical simulations is the range of scales where results are converged. In order to quantify this, we have used three simulations, all within the fiducial cosmology, but run at different resolutions: high-resolution ($1024^3$ particles), fiducial-resolution ($512^3$ particles), and low-resolution ($256^3$ particles).

In Fig. \ref{fig:void_finder} we show the projected matter overdensity field in a slice of $500\times500\times10~(h^{-1}{\rm Mpc})^3$ for the three different simulations. As the amplitudes and phases of the modes that are common across the simulations are the same, the large-scale density field in the three images is basically the same. Differences show up on small scales, where different modes across simulations are present/absent. Resolution effects are clearly visible in the image: while in the low-resolution simulation we can see individual particles in cosmic voids, in the high-resolution the density field is much smoother.

\begin{figure}
\begin{center}
\includegraphics[width=0.49\textwidth]{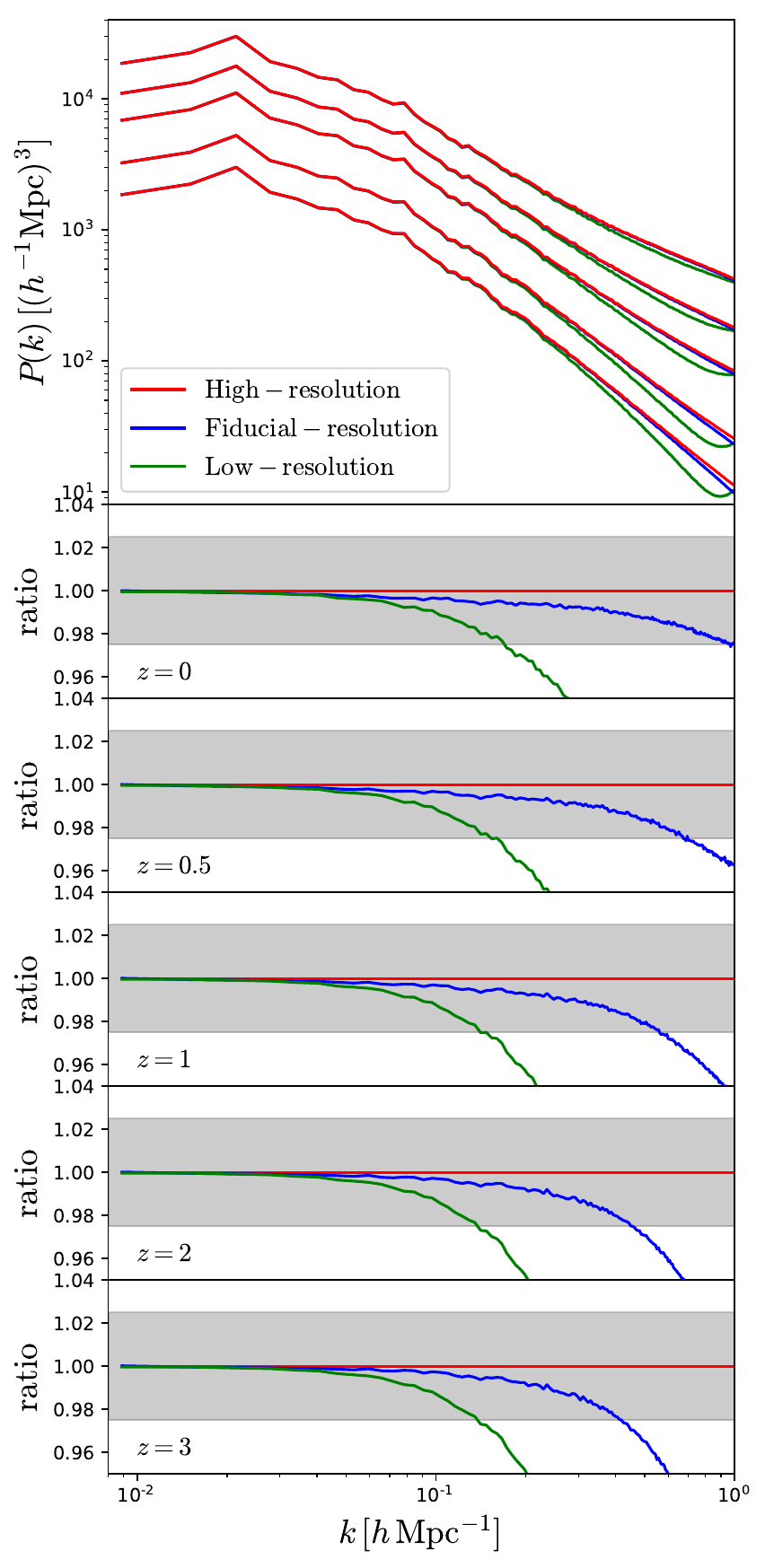}
\caption{Matter power spectrum for a realization of the fiducial cosmology run at three different resolutions: 1) high-resolution (red lines), 2) fiducial-resolution (blue lines), and 3) low-resolution (green lines). The upper panel shows the different power spectra from $z=0$ (top-lines) to $z=3$ (bottom lines). The small panels display the ratio between the different power spectra at different redshifts. At $z=0$, the matter power spectrum is converged all the way to $k=1~h{\rm Mpc}^{-1}$ at $2.5\%$, while at $z=3$ this scale shrinks to $k\simeq0.4~h{\rm Mpc}^{-1}$. }
\label{fig:resolution_test_Pk}
\end{center}
\end{figure}

We have computed the matter power spectrum for those three simulations at redshifts 0, 0.5, 1, 2, and 3. We show the results in Fig. \ref{fig:resolution_test_Pk}. We find that at $z=0$, the results of the fiducial-resolution run are converged all the way to $k=1~h{\rm Mpc}^{-1}$ at $2.5\%$. At higher redshifts, the results are only converged on larger scales; e.g. at $z=3$, only scales $k\simeq0.4~h{\rm Mpc}^{-1}$ are converged at the fiducial-resolution. We note that although the relative small scales error increases with redshift on, the absolute error do decrease, since the amplitude of the power spectrum shrinks with redshift.

We emphasize that these tests indicate the range of scales where the absolute amplitude of the clustering should be trusted within a given accuracy. Numerical derivatives of statistics with respect to cosmological parameters may be converged to smaller scales, since it is expected that relative differences propagate among models in a systematic manner, such as taking differences will cancel  the systematic bias.

\subsection{Void finder}
\label{subsec:VF_resolution}

The void finder (see subsection \ref{subsec:void_catalogues}) we have run on the \textsc{Quijote} simulations has some nice properties. One of them, is that the positions and sizes of cosmic voids are not largely affected by the mass and spatial resolution of the simulation\footnote{We are of course assuming that the sizes of the voids are larger than the spatial resolution of the density field.}.

In Fig. \ref{fig:void_finder} we show the location and sizes of voids identified in 3 different simulations with the same random seed but different mass and spatial resolutions. As can be seen, the locations and sizes of voids among simulations are very similar, pointing out the robustness of our void finder against mass and spatial resolution.

\section{Summary}
\label{sec:Conclusions}

In this paper we have introduced the \textsc{Quijote} simulations, a large set of 44100 full N-body simulations spanning thousands of different cosmologies and containing, at a single redshift, more than 8.5 trillion ($8.5\times10^{12}$) particles. Each simulation follows the evolution of $256^3$ (low-resolution), $512^3$ (fiducial-resolution) or $1024^3$ (high-resolution) CDM particles in a periodic volume of $1~(h^{-1}{\rm Gpc})^3$.
Billions of dark matter halos and cosmic voids have been identified in the simulations, that required more than 35 million CPU hours to be run.

The \textsc{Quijote} simulations have been designed to accomplish two main goals
\begin{itemize}
\item Quantify the information content on cosmological observables
\item Provide enough statistics to train machine learning algorithms
\end{itemize}

It is clear that there are many possible uses for these simulations beyond the ones we have mentioned in here \citep[see e.g.][]{Andrej_19}. We  make the data from the \textsc{Quijote} simulations freely available to the community with the goal to allow the broadest possible exploration of their applications.

We believe the \textsc{Quijote} simulations will complement very well the large efforts carried out by the community \citep[see e.g.][]{Aemulus, Nishimichi_2018, Coyote,EuclidEmu,Abacus}.

Instructions on how to download the data can be found in \url{https://github.com/franciscovillaescusa/Quijote-simulations}. As far as our storage resources allow, we will distribute all data products: e.g. halo and void catalogues, power spectra, marked power spectra, correlation functions, bispectra, PDFs, and full snapshots. The total data generated by the \textsc{Quijote} simulations exceeds 1 Petabyte.

We also provide a set of python libraries, \textsc{Pylians}, developed for many years, to help with the analysis of the data. \textsc{Pylians} can be found in \url{https://github.com/franciscovillaescusa/Pylians}.

\section*{ACKNOWLEDGEMENTS}
We are specially thankful to Nick Carriero and Dylan Simon from the Flatiron Institute, and Mahidhar Tatineni from the San Diego Supercomputer Center for their immense help with the multiple technical problems we have faced while running the simulations. We thank Volker Springel for giving us access to Gadget-III. The work of SH,  EG, EM, DS, FVN, and BDW  is supported by the Simons Foundation. CDK acknowledges the support of the National Science Foundation award number DGE1656466 at Princeton University. AP is supported by NASA grant 15-WFIRST15-0008 to the WFIRST Science Investigation Team ``Cosmology with the High Latitude Survey''. LV acknowledges support from the European Union Horizon 2020 research and innovation program ERC (BePreSySe, grant agreement 725327) and MDM-2014-0369 of ICCUB (Unidad de Excelencia Maria de Maeztu. TC and BDW also acknowledge financial support from ANR
BIG4 project, under reference ANR-16-CE23-0002. AMD acknowledges support from AstroCom NYC, NSF award AST-1831412 and Simons Foundation award number 533845. SH thanks NASA for their support in grant number: NASA grant 15-WFIRST15-0008 and NASA Research Opportunities in Space and Earth Sciences grant 12-EUCLID12-0004.

\bibliography{references}{}
\bibliographystyle{hapj}

\end{document}